\documentclass[fleqn,usenatbib]{mnras}

\usepackage{newtxtext,newtxmath}
\usepackage{comment}
\usepackage{multirow}
\usepackage{subcaption}
\usepackage{longtable}
\usepackage{supertabular}
\usepackage{anyfontsize}
\usepackage{placeins}

\usepackage[T1]{fontenc}
\usepackage{booktabs}
\usepackage{siunitx}

\DeclareRobustCommand{\VAN}[3]{#2}
\let\VANthebibliography\thebibliography
\def\thebibliography{\DeclareRobustCommand{\VAN}[3]{##3}\VANthebibliography}

\usepackage{graphicx}	
\usepackage{amsmath}
\usepackage{amsbsy}
\usepackage{subcaption}
\usepackage{balance}

\title[Merger-Triggered Star Formation and AGN]{Galaxy Mergers in the Epoch of Reionization II: Major Merger-Triggered Star Formation and AGN Activities at $z = 4.5 - 8.5$}

\author[Qiao Duan et al.]{
Qiao Duan,$^{1,2,3}$ Qiong Li,$^{1}$\thanks{E-mail: qiong.li@manchester.ac.uk} Christopher J. Conselice,$^{1}$\thanks{E-mail: conselice@manchester.ac.uk} Thomas Harvey,$^{1}$\thanks{E-mail: thomas.harvey-3@manchester.ac.uk} Duncan Austin,$^{1}$ Nathan J. Adams,$^{1}$ \and Leonardo Ferreira,$^{4}$ Kenneth J. Duncan,$^{5}$ James Trussler,$^{1,6}$ Robert G. Pascalau,$^{2,3}$ Rogier A. Windhorst,$^{7}$ \newauthor  Benne W. Holwerda,$^{8}$ Thomas J. Broadhurst,$^{9,10,11}$ Dan Coe,$^{12,13,14}$ Seth H. Cohen,$^{7}$ Xiaojing Du,$^{15}$ \newauthor Simon P. Driver,$^{16}$ Brenda Frye,$^{17}$ Norman A. Grogin,$^{12}$ Nimish P. Hathi,$^{12}$ Rolf A. Jansen,$^{7}$ \newauthor  Anton M. Koekemoer,$^{12}$ Madeline A. Marshall,$^{18,19}$ Mario Nonino,$^{20}$ Rafael {Ortiz~III},$^{7}$ Nor Pirzkal,$^{12}$ \newauthor Aaron Robotham,$^{16}$  Russell E. Ryan, Jr$^{12}$   Jake Summers,$^{7}$  Jordan C. J. D'Silva,$^{16,20}$ Christopher N. A. Willmer,$^{17}$ \newauthor Haojing Yan$^{21}$  
\\
Affiliations are listed at the end of the paper
}

\pubyear{2024}

\begin{document}
\label{firstpage}
\pagerange{\pageref{firstpage}--\pageref{lastpage}}
\maketitle

\begin{abstract}
Galaxy mergers are a key driver of galaxy formation and evolution, including the triggering of AGN and star formation to a still unknown degree. We thus investigate the impact of galaxy mergers on star formation and AGN activity using a sample of 3,330 galaxies at $z = [4.5, 8.5]$ from eight JWST fields (CEERS, JADES GOODS-S, NEP-TDF, NGDEEP, GLASS, El-Gordo, SMACS-0723, and MACS-0416), collectively covering an unmasked area of 189 arcmin$^2$. We focuses on star formation rate (SFR) enhancement, AGN fraction, and AGN excess in major merger ($\mu > 1/4$) close-pair samples, defined by $\Delta z < 0.3$ and projected separations $r_p < 100$ kpc, compared to non-merger samples. We find that SFR enhancement occurs only at $r_p < 20$ kpc, with values of $0.25 \pm 0.10$ dex and $0.26 \pm 0.11$ dex above the non-merger medians for $z = [4.5, 6.5]$ and $z = [6.5, 8.5]$, respectively. No other statistically significant enhancements in galaxy sSFR or stellar mass are observed at any projected separation or redshift bin. We also compare our observational results with predictions from the SC-SAM simulation and find no evidence of star formation enhancement in the simulations at any separation range. Finally, we examine the AGN fraction and AGN excess, finding that the fraction of AGNs in AGN-galaxy pairs, relative to the total AGN population, is \(3.25^{+1.50}_{-1.06}\) times greater than the fraction of galaxy pairs relative to the overall galaxy population at the same redshift. We find that nearly all AGNs have a companion within 100 kpc and observe an excess AGN fraction in close-pair samples compared to non-merger samples. This excess is found to be \(1.26 \pm 0.06\) and \(1.34 \pm 0.06\) for AGNs identified via the inferred BPT diagram and photometric SED selection, respectively.
\end{abstract}

\section{Introduction}
Galaxy mergers are a fundamental mechanism driving the evolution of galaxies. Theory and simulations show that tidal interactions from mergers induce bar instabilities in the centers of galaxies, leading to strong gravitational torques that funnel gas toward the central regions, triggering both starbursts and active galactic nucleus (AGN) activity \citep[e.g.,][]{1996ApJ...471..115B, 1996ApJ...464..641M, 2006MNRAS.373.1013C, di2007star, 2008ApJS..175..356H, montuori2010dilution, rupke2010galaxy, torrey2012metallicity, 2012MNRAS.426..549S, 2015MNRAS.448.1107M, 2019MNRAS.485.1320M}. 

Observational studies have provided crucial insights into how star formation evolves during mergers, but a comprehensive understanding of the temporal and environmental factors influencing these changes is still developing. In particular, there is a need to trace variations in the star formation rate (SFR) across different stages of mergers and at various cosmic epochs to quantify the impact of these interactions on galactic evolution. Studies at low redshifts ($0 < z < 1$) have shown significant SFR enhancements in galaxy pairs. Using data from the Sloan Digital Sky Survey, multiple studies \citep[e.g.,][]{ellison2008galaxy,10.1111/j.1365-2966.2010.17076.x, 10.1111/j.1365-2966.2010.17932.x, 2012MNRAS.426..549S, 2013MNRAS.433L..59P, 2016MNRAS.461.2589P} have reported that galaxies in close pairs (with projected separations of less than 40 - 80 kpc) exhibit, on average, a 60\% higher SFR compared to controlled isolated galaxies. Similarly, \cite{2021MNRAS.501.2969G}, using data from the Galaxy and Mass Assembly (GAMA) survey \citep{2011MNRAS.413..971D}, found SFR enhancements ranging from $0.025$ to $0.15$ dex. At higher redshifts, \cite{2014AJ....148..137L} identify merging galaxies in the COSMOS field between $0.25 < z < 1.00$ at projected separations of $r_p = $ 2.2 - 8 kpc, reporting an SFR enhancement of $2.1 \pm 0.6$ in their merging sample compared to non-merging galaxies. Similarly, \cite{2022ApJ...940....4S} found an SFR enhancement factor of $\sim 1.23^{+0.08}_{-0.09}$ in the redshift range $0.5 < z < 3$. Besides the aforementioned literature, there are several works who have verified the SFR enhancement in close-pairs, e.g., \cite{2015MNRAS.452..616D, 2021MNRAS.503.3113M, 2024arXiv240913014C}.
2
However, there are studies which have found the opposite of this.  In contrast to induced star formation, \cite{2018ApJ...868...46S} find no significant difference in star formation properties between merger and non-merger galaxies at $0.5 < z < 2.5$ in the projection separation range of $3 < r_p < 15$. \cite{2019A&A...631A..51P}, using convolutional neural networks to detect binary mergers in SDSS (DR7), KiDS \citep{de_Jong_2012}, and CANDELS \citep{2011ApJS..197...35G, 2011ApJS..197...36K} imaging, reported only a 20\% increase in SFR due to merger activity. Additionally, \cite{ellison2022galaxy} discovered that the frequency of post-merger galaxies that rapidly shut down their star formation after a starburst is 30 to 60 times higher than that of a control sample of non-merging galaxies. Interestingly, no such excess shutdown was found among close galaxy pairs, suggesting that while mergers can trigger star formation, they can also lead to its rapid cessation, particularly after the merging process is complete. \cite{2023ApJ...953...91L}, analyzing SDSS data within the redshift range $0.01 < z < 0.11$, find that in low-redshift galaxies, mergers—including those between gas-rich and nearly equal-mass galaxies—have a negligible effect on their SFR, specific SFR, and star formation efficiency. Starbursts remain uncommon, and the star formation efficiency of gas-rich minor mergers may even be reduced.

There are two primary observational approaches to investigate the impact of mergers on galaxy star forming properties. One approach uses morphological parameters, deep learning, and convolutional neural network methods on galaxy morphology to identify merging galaxies and compare their star-forming properties with controlled non-merging galaxies \citep[e.g.,][]{2014AJ....148..137L, 2018ApJ...868...46S, 2019A&A...631A..51P, dalmasso2024rate, 2025MNRAS.538L..31F}. The other approach is the close-pair method, which investigates the extent of SFR enhancement at various projected separations to the closest neighbor \citep[e.g.,][]{10.1111/j.1365-2966.2010.17076.x,2012MNRAS.426..549S, 2022ApJ...940....4S}. The morphological method is influenced by various intrinsic and extrinsic galaxy properties, such as orbit type, gas fraction, viewing angle, wavelength, redshift, and depth \citep{2011ApJ...742..103L, 2020MNRAS.492.2075B}, which introduces challenges in identifying merging galaxies at the very high-redshifts ($z = [4.5, 8.5]$) used in this study. Therefore, we conduct an early JWST analysis using the close-pair method to investigate the enhancement of galaxy star-forming properties for the most distant galaxies.

Besides triggering star formation, galaxy mergers can also drive rapid gas accretion onto supermassive black holes (SMBHs), thereby initiating AGN activity \citep[e.g.,][]{Koulouridis_2006_1, Koulouridis_2006, ellison2011galaxy, koulouridis2013activity, koulouridis2014dichotomy,satyapal2014galaxy, 2017MNRAS.468.1273R, 2017ApJ...850...27B, 2020A&A...637A..94G, 2023MNRAS.522.1736P, 2023MNRAS.522.5165A, 2023MNRAS.519.4966B}. Inflowing material can obscure the central regions, making the most heavily obscured AGNs elusive to optical and soft X-ray surveys. Infrared observations offer a more reliable detection method due to their lower susceptibility to absorption \citep{2019PASJ...71...31C}. With the high resolution of the JWST NIRSpec instrument, AGNs can be identified either through the broad-line components of key emission lines (H$\alpha$, H$\beta$) \citep{2023ApJ...959...39H, 2024arXiv240500504M, 2024Natur.627...59M} or using diagnostic diagrams \citep{2001ApJ...556..121K, 2003MNRAS.346.1055K, 2006MNRAS.372..961K, 2023A&A...679A..89P, 2023arXiv231118731S, 2024arXiv240815615M}. Photometrically, AGN detection can also be inferred from SED templates \citep{2022MNRAS.513.5134N, 2023MNRAS.525.1353J}.

Understanding the fraction of galaxies that harbor AGNs is crucial for grasping the black hole accretion history of the Universe. \cite{2024ApJ...963...53C} analyzed MaNGA data and identified an AGN excess of about 1.8 in major mergers and 1.5 in minor mergers relative to non-merger isolated samples. \cite{2020A&A...637A..94G} investigate AGN fractions using low-redshift SDSS samples at $0.005 < z < 0.1$ \citep{2009ApJS..182..543A} and slightly higher redshift samples from GAMA ($0 < z < 0.6$) \citep{2011MNRAS.413..971D, 2015MNRAS.452.2087L}. Their study revealed a higher AGN fraction in merger samples compared to non-merger samples. Specifically, they report AGN fractions of 1.63 ± 0.05\% versus 1.45 ± 0.02\% in optical and 0.34 ± 0.02\% versus 0.20 ± 0.01\% in mid-infrared (MIR) for SDSS AGNs and controls, respectively. Similarly, in GAMA, the optical AGN fractions are found to be 3.63 ± 0.11\% for mergers versus 2.53 ± 0.03\% for the controls; within the MIR the fractions are 0.73 ± 0.05\% versus 0.53 ± 0.01\%. In addition, numerous studies have shown that the AGN fraction increases as the separation between galaxy pairs decreases \citep{ellison2011galaxy, silverman2011impact, koss2012understanding, ellison2013galaxy, satyapal2014galaxy, khabiboulline2014changing,bickley2023, 2024MNRAS.533.3068B}.

Most of the literature on merger triggered star formation and AGN activity focuses on redshifts $z < 4.5$. The process at $z > 4.5$ epoch remains largely, if not completely, unknown. The high-redshift Universe is increasingly unveiled by JWST, highlighted by numerous recent studies on early galaxy discoveries \citep[e.g.,][]{2022ApJS..259...20H, 2022MNRAS.511.4464A, 2022ApJ...938L..15C, 2022NatAs...6..599D, 2022ApJ...940L..14N, curtis2022spectroscopic, 2023MNRAS.518.4755A, 2023arXiv230801230M,2023ApJ...952L...7A,donnan2023evolution, finkelstein2023ceers, 2023Natur.622..707A, 2023MNRAS.523.1036B, 2023arXiv230414469M, 2023arXiv230602468H, 2023arXiv230800751F, 2023arXiv230810932C, 2024Natur.627...59M,2024ApJ...965..169A, 2024arXiv240605306C, 2024arXiv240714973C}. These studies reveal a significant population of distant galaxies at $z > 4.5$, previously underrepresented in HST data, providing a new window to explore merger evolution at high redshifts. This paper is the second in the \textit{Galaxy Mergers in the Epoch of Reionization} series. The first paper in this series \citep{2024arXiv240709472D} presented, for the first time, galaxy pair fractions, merger rates, and stellar mass accretion at $z = 4.5 - 11.5$. It revealed a rapid merger rate of approximately 6 mergers per Gyr at these redshifts, with mergers contributing $71 \pm 25\%$ of the stellar mass growth in galaxies, comparable to the increase from star formation fueled by gas. In this paper, we extend this work to explore the impact of mergers on galaxy star formation and AGN activity at $z = 4.5 - 8.5$, using the same data as the first study from eight JWST observational fields: CEERS, JADES GOODS-S, NEP-TDF, NGDEEP, GLASS, El-Gordo, SMACS-0723, and MACS-0416.

The structure of this paper is as follows. In Section \ref{sec: Observations and Data Reduction}, we describe the data we use and outline our methods for redshift measurement, star formation rate and stellar mass inference. Our approach for classifying close-pair galaxies is detailed in Section \ref{sec: close-pair classifications}. The main findings are presented in Section \ref{sec: results}. Finally, we discuss our results in Section \ref{sec: Discussion} and summarize our conclusions in Section \ref{sec: Conclusions}. Throughout this work, we adopt the Planck 2018 cosmology \citep{2020A&A...641A...6P}, with $H_0 = 67.4 \pm 0.5\,\text{km}\,\text{s}^{-1}\,\text{Mpc}^{-1}$, $\Omega_{\rm M} = 0.315 \pm 0.007$, and $\Omega_{\Lambda} = 0.685 \pm 0.007$, to ensure consistency with other observational studies. All magnitudes reported are in the AB magnitude system \citep{Oke1974,Oke1983}.

\section{Observations, Data Reduction and Data Products}
\label{sec: Observations and Data Reduction}
The launch of the James Webb Space Telescope (JWST) in December 2021 \citep{2023PASP..135d8001R} has ushered in a new era of high-redshift Universe exploration. In this work, we consistently reduce and analyze publicly available deep JWST NIRCam data from the PEARLS, CEERS, GLASS, JADES GOODS-S, NGDEEP, and SMACS0723 surveys. Specifically, for the PEARLS program, we include data from the NEP-TDF, MACS-0416, and El-Gordo fields, resulting in a combined total of 189arcmin$^2$ of unmasked sky when integrated with the other five surveys. From this extensive dataset, we have identified 2,404 robust galaxy candidates at $z = [4.5, 6.5]$, with 1,276 candidates at $z > 6.5$ (926 at $z = [6.5, 8.5]$), forming our EPOCHS V1 sample \citep[][]{conselice2024epochs}. For the \(z = 4.5 - 6.5\) range, we utilize data only from the CEERS, JADES GOODS-S, and NEP-TDF surveys due to the availability of HST ACS/WFC F606W and F814W band data.

For the SMACS-0723, MACS-0416, and El-Gordo cluster observations, one NIRCam module is centered on the lensing cluster in each pointing, while the other module is positioned approximately 3 arcminutes away, providing an effectively "blank-field" view of the distant Universe. In this paper, we exclude sources found in the cluster module to avoid potential uncertainties in corrections for mergers and contamination from cluster members, as well as complications arising from lensing effects \citep[e.g.,][]{Griffiths2018, Bhatawdekar2021}. For these clusters, only the NIRCam module that is not centered on the cluster is used.

We utilize our own EPOCHS reduction pipeline, with details provided in \cite{2024ApJ...965..169A, austin2024epochs, harvey2024epochs, conselice2024epochs}. While we do not repeat the full details here, the methods are thoroughly discussed in these papers. It is essential to note that all fields are reprocessed using the same methodology, ensuring uniformity in galaxy detection and measurement. We have reprocessed all lower-level JWST data products using a modified version of the official JWST pipeline. For further details on the data reduction process and catalogs, please refer to \cite{2024ApJ...965..169A, austin2024epochs, harvey2024epochs, conselice2024epochs}.

In the following subsections, we describe the various fields studied and the relevant properties for this paper. The depth of our data, collected in various fields and JWST filters, as well as in HST ACS, is outlined in Table 1 of \cite{2024ApJ...965..169A} within \citet{2024arXiv240714973C}.

\subsection{CEERS}

The Cosmic Evolution Early Release Science Survey \citep[CEERS; PI: S. Finkelstein, PID: 1345,][]{2023ApJ...946L..12B} is one of the 13  Early Release Science (ERS) programs conducted during the first year of JWST's operation. CEERS features a $\sim$100 arcmin$^2$ NIRCam imaging mosaic covering wavelengths from 1 to 5 microns, across 10 NIRCam pointings. Concurrently, CEERS includes imaging with the mid-infrared instrument (MIRI) over eight pointings, covering 5 to 21 microns, and multi-slit spectroscopy using the near-infrared spectrograph (NIRSpec) across six pointings. NIRCam and HST imaging data is used in this work. This includes imaging across nine distinct filters: F606W, F814W, F115W, F150W, F200W, F277W, F356W, F410M, and F444W, with a $5\sigma$ depth of 28.6 AB magnitudes using 0.16 arcsec radius circular apertures. 

In addition to photometric data, we also use CEERS Spectroscopic data. The CEERS NIRSpec spectroscopic data \citep{fujimoto2023ceers, 2023Natur.622..707A,2023ApJ...951L..22A} were procured as part of the ERS program (PI: Steven L. Finkelstein, ID:1345). This dataset was designed to optimize the overlap with observations from both NIRCam and HST, using three medium resolution gratings \( R \approx 1000 \) and the PRISM \( R \approx 100 \). During this period, both NIRSpec pointings, namely NIRSpec11 and NIRSpec12, adhered to the standard CEERS MSA observational guidelines. Specifically, they encompassed three integrations with 14 groups in the NRSIRS2 readout mode per visit, leading to a total exposure time of 3107 s. Within these observations a trio of shutters was used to form slitlets, facilitating a three-point nodding sequence to enhance background subtraction. The PRISM disperser, ranging in wavelength from 0.6–5.3 $\mu$m, is characterized by its capacity to provide varied spectral details. In this paper, we use the NIRSpec data reduced by the Cosmic Dawn Center, which is published on the DAWN JWST Archive (DJA).\footnote{\href{https://dawn-cph.github.io/dja/blog/2023/07/18/nirspec-data-products/s p}{https://dawn-cph.github.io/dja/blog/2023/07/18/nirspec-data-products/}.}

\subsection{JADES Deep GOODS-S}
The JWST Advanced Deep Extragalactic Survey \citep[JADES;][]{rieke2023jades, 2023arXiv230602465E, bunker2023jades, 2024arXiv240406531D, 2023AAS...24221202H} cover both the GOODS-S and GOODS-N fields. In this paper, we focus on the region covered by the JADES DR1 release which is in the GOODS-S field (PI: Eisenstein, N. Lützgendorf, ID:1180, 1210). The observations utilise nine filter bands: F090W, F115W, F150W, F200W, F277W, F335M, F356W, F410M, and F444W, encompassing a spatial extent of 24.4 - 25.8 arcmin\(^2\). A minimum of six dither points was used for each observation, with exposure times spanning 14-60 ks. The depth of the JADES data within the JWST bands ranges from 29.58 to 30.21, with the deepest band F277W. We also use the HST reductions from v2.5 of the Hubble Legacy Fields Project \citep{2016arXiv160600841I, 2019ApJS..244...16W}, including ACS/WFC3 F606W and F814W filterband data.

We also incorporate public JADES NIRSpec data. We use the third JADES released NIRSpec \citep{ferruit2022near, bunker2023jades, 2024arXiv240406531D} data (PI: Eisenstein, N. Lützgendorf, ID:1180, 1210), with a focus on the publicly released data in GOODS-S field. The spectra are obtained through the application of both disperser/filter and PRISM/clear configurations. Four different disperser/filter combinations are used to acquire the spectroscopy: G140M/F070LP, G235M/F170LP, G395M/F290LP, and G395H/F290LP, with a wavelength coverage of $0.70 - 1.27 \mu$m, $1.66 - 3.07 \mu$m, $2.87 - 5.10 \mu$m, and $2.87 - 5.14 \mu$m, respectively. The three medium resolution filters have a nominal resolving power of R $\approx 1000$, while the high resolution data can reach R $\approx 2700$. In this paper, we primarily utilize the PRISM data, which covers a wavelength range of \(0.6 \, \mu\text{m}\) to \(5.3 \, \mu\text{m}\), and exhibits a spectral resolution of \( R \approx 30 - 330\) \citep{ji2022reconstructing}.

\subsection{PEARLS and NEP-TDF}
\label{sec: NEP NIRCam}
The North Ecliptic Pole Time-Domain Fields (NEP-TDF) is part of the JWST PEARLS observational program. The Prime Extragalactic Areas for Reionization and Lensing Science project \citep[PEARLS; PIDs 1176, 2738, PI: Rogier Windhorst;][]{2023AJ....165...13W,2023A&A...672A...3D, 2023ApJ...952...81F}, is a JWST Guaranteed Time Observation (GTO) program. As a Cycle 1 GTO project, the PEARLS team was allocated a science time of 110 hours. The project's primary objective is to capture medium-deep NIRCam imaging of blank and cluster fields with an approximate depth of 28–29 AB magnitudes. 

In this study, we use the first two lensing clusters observed by PEARLS, MACS-0416 and El Gordo, along with NEP-TDF. The data acquired from PEARLS includes imaging in seven wide filter bands: F090W, F115W, F150W, F200W, F277W, F356W, and F444W, along with one medium band: F410M. In addition, we also include bluer F606W imaging from the HST Advanced Camera for Surveys (ACS) Wide Field Channel (WFC) in the NEP-TDF from the GO-15278 (PI: R.~Jansen) and GO-16252/16793 \citep[PIs: R.~Jansen \& N.~Grogin, see][]{2024ApJS..272...19O} HST programs.

For the lensing fields MACS 0416, and El Gordo, the observations include pointings with one NIRCam module centered on the lensing cluster, while the second module is offset by around 3 arcminutes in a 'blank' region. Although we process both modules in these fields, we do not include sources found in the module containing the lensing cluster in this study.



\subsection{NGDEEP}
The Next Generation Deep Extragalactic Exploratory Public Survey (NGDEEP; ID 2079, PIs: S. Finkelstein, Papovich, and Pirzkal) \citep{2023ApJ...952L...7A, 2023arXiv230205466B, 2023ApJ...954L..46L} is the deepest public survey with JWST NIRCam conducted in the first year of observations. Initially scheduled for late January to early February 2023, NGDEEP's primary focus involved NIRISS Wide Field Slitless Spectroscopy of galaxies within the Hubble UltraDeep Field. However, due to an unforeseen suspension of NIRISS operations during the survey's observation window, only half of the planned observations were executed, with the remainder taken in early 2024. In this work we use exposures from the second epoch of observations. The parallel NIRCam observations from NGDEEP encompass six wide-band photometric filters (F115W, F150W, F200W, F277W, F356W, F444W). In addition, we also use HST data from the same Hubble Legacy Fields project as JADES. Some of our early work in this field was presented in \citet{austin2023large}.

\subsection{GLASS}
The Grism Lens Amplified Survey from Space (GLASS) observational program \citep[GLASS; PID 1324, PI: T. Treu;][]{2022ApJ...935..110T} is centered on the Abell 2744 Hubble Frontier Field lensing cluster. GLASS primarily employs NIRISS and NIRSpec spectroscopy to study galaxies lensed by the Abell 2744 cluster. In parallel, NIRCam is used to observe two fields offset from the cluster center. Due to the minimal lensing magnification in these offset areas \citep{2023A&A...670A..60B}, they effectively serve as blank fields. The observations include seven wide filter bands: F090W, F115W, F150W, F200W, F277W, F356W, and F444W.

\subsection{SMACS J0723-7327 ERO}
The SMACS-0723 lensing cluster is part of the very first ERO program (PID 2736, PI: Klaus Pontoppidan, \cite{2022ApJ...936L..14P}). The program includes NIRCam, MIRI, NIRISS, and NIRSpec observations, and we use the NIRCam observations which contains six bands: F090W, F150W, F200W, F277W, F356W, and F444W. The exposure time of this observation is set to roughly reach the Frontier depth in F814W, 1.5 times the F160W depth, and almost 10 times the Spitzer \citep{Werner2004} IRAC 1+2 depth.  Because of the inaccuracies of the photometric redshifts for galaxies within this field around $z \sim 10$ due to the lack of the F115W filterband, for the most part the pairs in this field are at the lower or higher redshifts of our range.

\subsection{Photometric Redshift Measures}
The photometric redshift measurements and selection criteria for our EPOCHS samples have been discussed in detail in the EPOCHS series papers. In this paper, we summarize the method. For an in-depth explanation, we refer readers to \cite{2024ApJ...965..169A}, \cite{austin2024epochs}, \cite{harvey2024epochs}, and \cite{conselice2024epochs}.

We use \texttt{EAZY-PY} (hereafter \texttt{EAZY} \citep{brammer2008eazy}) as our primary photometric SED-fitting code. We employ the "$\mathrm{tweak\_fsps\_QSF\_12\_v3}$" templates alongside Sets 1 and 4 of the SED templates developed by \citet{2023ApJ...958..141L}. These additional templates are optimized for high-redshift galaxies, which exhibit bluer rest-frame UV colors \citep{2022ApJ...941..153T, 2023MNRAS.520...14C, 2023ApJ...947L..26N, 2024MNRAS.531..997C, austin2024epochs} and stronger emission lines \citep[e.g.][]{2023ApJ...958L..14W}, characteristics frequently observed in such young and distant galaxies.

The "$\mathrm{tweak\_fsps\_QSF\_12\_v3}$" templates are generated using the flexible stellar population synthesis (FSPS) package \citep{2010ascl.soft10043C}. Set 1 includes three pure stellar templates from \citet{2023ApJ...958..141L}, which are based on the BPASS v2.2.1 stellar population synthesis (SPS) model \citep{2017PASA...34...58E, 2018MNRAS.479...75S, 2022MNRAS.512.5329B} for ages of $10^6$, $10^{6.5}$, and $10^7$ years at a fixed metallicity $Z_\star = 0.05 \, \mathrm{Z}_\odot$, assuming a standard Chabrier \citep{2003PASP..115..763C} IMF. Set 4 incorporates nebular continuum and line emission calculated using CLOUDY v17 \citep{2017RMxAA..53..385F}, with an ionization parameter $\log U = -2$, gas metallicity $Z_{gas} = Z_\star$, Lyman continuum escape fraction $f_\mathrm{esc,LyC} = 0$, hydrogen density $n_\mathrm{H} = 300$ cm$^{-3}$, and a spherical geometry, excluding Ly$\alpha$ emission.

We execute the \texttt{EAZY-py} fitting three times: first with a uniform redshift prior of $0 \leq z \leq 25$, and subsequently with a constrained upper redshift limit of $z \leq 4$ and $z \leq 6$. This triple approach allows us to compare the goodness of fit between high and low-redshift solutions, as indicated by the $\chi^2$ values, ensuring robust identification of high-redshift galaxies.

To account for uncertainties in flux calibrations and aperture corrections, as well as potential mismatches between synthetic templates and observed galaxies, we impose a minimum flux uncertainty of 10\% \citep{2023PASP..135b8001R}. Our selection criteria rely predominantly on specific cuts in computed quantities to minimize biases and incompleteness. Although we perform a visual review of cutouts and SED-fitting  solutions, this step results in less than 5\% rejection of our original sample.

\subsection{Star Formation Rate and Stellar Mass Measures}
\label{sec: Stellar Mass Estimation}
To infer galaxy stellar masses from photometry, we use the \texttt{Bagpipes} \citep{carnall2018inferring} Bayesian SED fitting code, adopting redshift probability distribution functions from \texttt{EAZY} for each galaxy. We use logarithmic priors for dust, metallicity, and age. The reason for selecting logarithmic priors is because we expect high redshift galaxies to be young, with a lower metallicity and to contain little dust. We set prior limits for metallicity in the range of $[5\text{e-03}, 5] \, \text{Z}_{\odot}$, and we use a dust prior in the range of $[0.0001, 10.0]$ in $A_\text{V}$, ionization parameter $\mathrm{Log}_{10}(\mathrm{U})$ in the range of $[-4, -1]$. Bagpipes does not allow SFH to extend beyond $\tau_\text{U}$, the age of the Universe at that redshift. In addition, \cite{kroupa2001MNRAS.322..231K} IMF, \cite{2003MNRAS.344.1000B} SPS model, and the  \cite{calzetti2000dust} dust attenuation model is implemented \citep{2022MNRAS.510.5088B}.

We fit the galaxy photometry with six parametric SFH models — log-normal, delayed, constant, exponential, double delayed, and delayed burst—along with a non-parametric Continuity model \citep{2019ApJ...876....3L}. These models have been used in various works \citep[e.g.][]{2023Natur.619..716C,2023MNRAS.524.2312E,2023MNRAS.522.6236T, 2023MNRAS.519.5859W}. Different SFHs can have a large impact on inferred stellar mass and SFR \citep{furtak2021robustly, 2022ApJ...927..170T,  harvey2024epochs, wang2024quantifying}. We have chosen to present results inferred from our fiducial model, the log-normal SFH with 10Myr averaged timescale, because we expect high-redshift galaxies to exhibit rising or bursting star formation \citep{2024MNRAS.529.4728D}. Please refer to \citet{harvey2024epochs} for a thorough discussion of the topic of priors in SED fitting. 

\begin{figure}
    \centering
    \includegraphics[width=\linewidth]{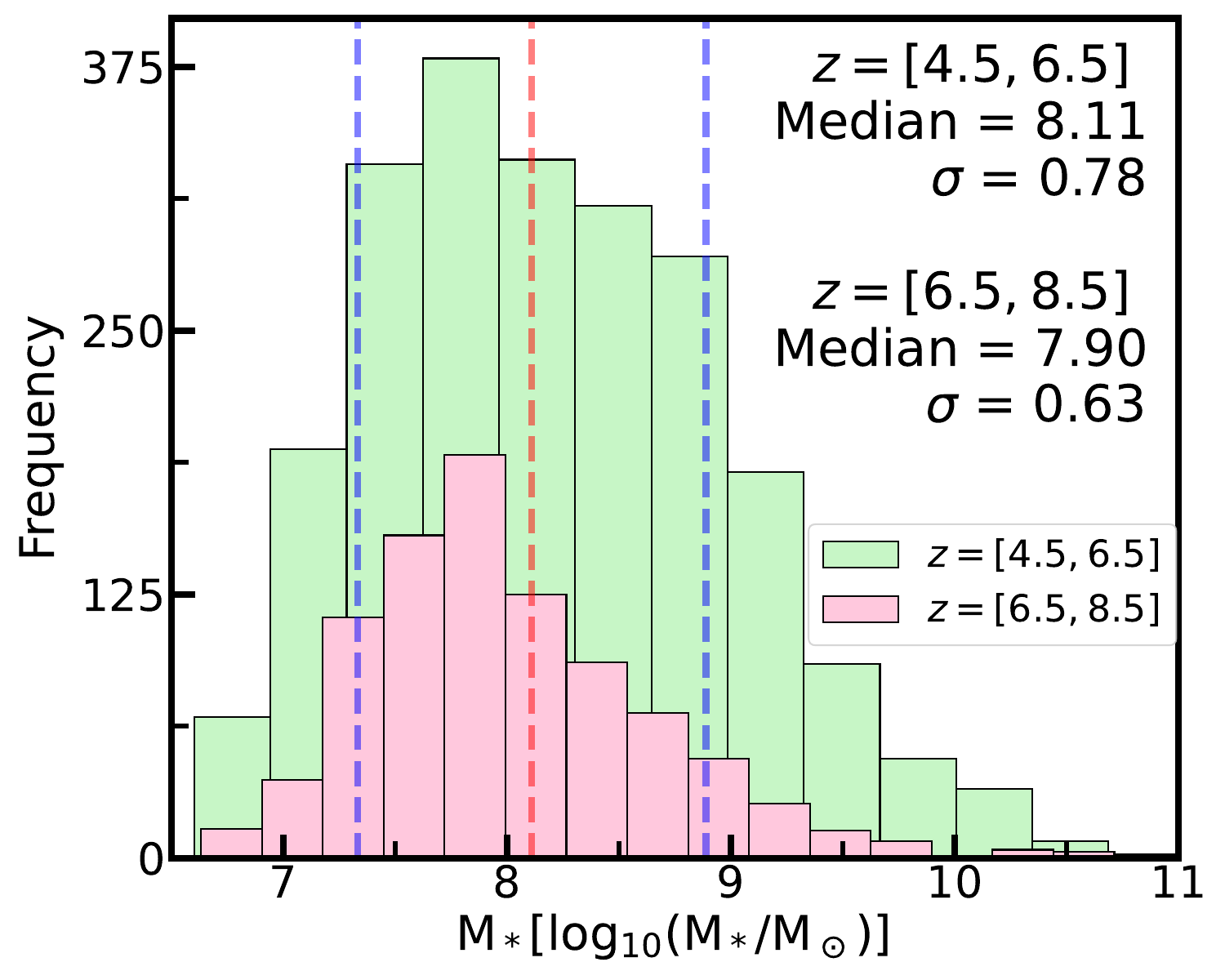}
    \caption{Histogram displaying the distribution of the best-fit stellar masses for all 2404 galaxies at $z = [4.5, 6.5]$ and for 926 galaxies at $z = [6.5, 8.5]$. These samples are from our eight  JWST fields: CEERS, JADES GOODS-S, NEP-TDF, NGDEEP, GLASS, El-Gordo, SMACS-0723, and MACS-0416. The median stellar masses are found to be $\sim 8.1$[$\mathrm{log}_{10}(\mathrm{M}_* / \mathrm{M}_\odot)$] and 7.9 [$\mathrm{log}_{10}(\mathrm{M}_* / \mathrm{M}_\odot)$] for each redshift bin respectively.  As can be seen, it is clear that there is a drop in completeness at $\mathrm{log}_{10} (\mathrm{M}_* / \mathrm{M}_\odot) < 8$ which is also seen in image simulations of these fields \citep[][]{harvey2024epochs}.}

    \label{fig: stellar mass distribution}
\end{figure}

We show the stellar mass distribution of 3300 galaxies used in this work in \autoref{fig: stellar mass distribution}. The green histogram represents the distribution of 2404 galaxies at $z = [4.5, 6.5]$, and the pink histogram shows the distribution of 926 galaxies at $z = [6.5, 8.5]$. We find the median stellar mass in these two redshift bins to be $\sim8$ $[\mathrm{log}_{10}(\mathrm{M}_* / \mathrm{M}_\odot)]$. To ensure as large a sample as possible, we use all galaxies within the mass range $\mathrm{log}_{10}(\mathrm{M}_* / \mathrm{M}_\odot) = [8,10]$ for later analysis.

\section{Close-Pair Classifications}
\label{sec: close-pair classifications}
We have two primary objectives: the first is to identify close-pair merging systems and analyze how Star Formation Rate (SFR), Specific Star Formation Rate (sSFR), and stellar mass change with proximity to the nearest companion. The second is to investigate AGN Fraction and AGN Excess in these systems. Having well-defined criteria for identifying close-pair samples is crucial, as it directly impacts the reliability of our results \citep{2024arXiv240709472D}. In this section, we outline the criteria used to classify galaxies into close-pair {\em major merger} and non-merger systems.

For a galaxy to be classified as part of the close-pair {\em major merger} sample, it must simultaneously satisfy the following three criteria. Galaxies that do not meet all three criteria are classified as non-merger samples. 
\begin{enumerate}
    \item \textbf{Projected Physical Separation} ($r_p$): The projected distance to the closest galaxy in the pair must be $r_p < 100$ kpc.
    \item \textbf{Redshift Difference} ($\Delta z$): The difference in redshift between the two galaxies must satisfy $\Delta z < 0.3$.
    \item \textbf{Mass Ratio} ($\mu$): The stellar mass of the lighter galaxy must be at least one-quarter of the stellar mass of the heavier galaxy, i.e., $\mu > 1/4$, for the pair to be classified as a major merger.  Typically, galaxies with lower mass ratios are classified as minor mergers. 
\end{enumerate}

The first criterion, projected physical separation, is widely used in previous studies of the local Universe with SDSS data \citep{ellison2008galaxy, 10.1111/j.1365-2966.2010.17932.x, 2012MNRAS.426..549S, 2013MNRAS.433L..59P}, as well as at higher redshifts \citep{2022ApJ...940....4S} ($0.5 < z < 3.0$). In our approach, we apply the KD Tree algorithm to calculate the projected distance to the closest neighbor for each galaxy. The mass ratio criterion for defining major mergers is also widely adopted in several studies \citep[e.g.,][]{2014MNRAS.445.1157C, 2015A&A...576A..53L, 2016ApJ...830...89M, 2017MNRAS.470.3507M, 2018MNRAS.475.1549M, 2019ApJ...876..110D, 2022ApJ...940..168C, 2024arXiv240709472D, 2025arXiv250201721P}. The key difference in the classification criteria used in this work is the second one — the Redshift Difference ($\Delta z$) criterion.

Various work at lower redshift ($z < 3$) apply a rest-frame velocity difference criteria instead of a redshift difference. For example, in the local Universe, \cite{ellison2008galaxy} apply a velocity difference limit of $\Delta v < 500$ km/s, while \cite{2012MNRAS.426..549S} apply a stricter limit of $\Delta v < 300$ km/s. At $0.5 < z < 3.0$, \cite{2022ApJ...940....4S} use both $\Delta v < 1000$ km/s and $\Delta v < 5000$ km/s as velocity difference limits. There are two main reasons for using $\Delta z$ in this work. Firstly, the previous papers all use spectroscopic observations, which enable highly accurate redshift measurements, whereas photometric redshifts are used in this work. Due to the faint nature of the high-redshift galaxies in our sample, spectroscopic observations require extended integration times of up to 28 hours. As a result, it is not possible to obtain spectroscopic observations for all galaxies detected photometrically. Additionally, the current number of available JWST spectroscopic observations are not statistically sufficient for this purpose. Secondly, the redshift range used in this work, $z = [4.5, 8.5]$, is wider than in all previous studies. Given the definition $\Delta v = \frac{c \times \Delta z}{(1 + z_{\mathrm{mean}})}$, the corresponding $\Delta v$ value for a given $\Delta z$ varies largely for pairs at different redshifts. Since the average photometric uncertainties in our sample are around $0.1$, we choose $\Delta z < 0.3$ as our fiducial limit for classifying close-pair galaxies. In the next section, we present our results using our fiducial $\Delta z$ criterion, and in \autoref{sec: appendix 1}, we show and discuss results under five different $\Delta z$ limits: 0.1, 0.2, 0.3, 0.4, and 0.5.

\section{Results}
\label{sec: results}
\subsection{Merger-Driven Star Formation Activity}
\label{sec: star_formation_enhancement}

In this section, we describe the methods used to investigate the enhancement of star formation in galaxy pairs at various separations and present the corresponding results. We then compare our observational findings with predictions from the \textit{Semi-analytic Forecasts for JWST} (SC-SAM) simulation, as detailed in Section \ref{sec: star_formation_enhancement_simulation}.

Using the close-pair classifications described in Section \ref{sec: close-pair classifications}, we classify galaxies into two groups. Those that satisfy all three close-pair criteria are placed in the close-pair group, while those that do not satisfy all three criteria are classified as the non-merger group. The non-merger group is used as a control set of galaxies to investigate the extent of star formation enhancement in close-pair galaxies. We then separate our redshift range into two bins: $z = [4.5, 6.5]$ and $z = [6.5, 8.5]$, and the enhancement in each redshift bin is investigated separately. To maintain sample completeness, we use all galaxy samples with masses in the range [8, 10] [$\mathrm{log}_{10}(\mathrm{M}_* / \mathrm{M}_\odot)$], as shown in \autoref{fig: stellar mass distribution}. After doing this we then plot the SFR as a function of $r_p$ and investigate the SFR enhancements at various separations. In total, 420 galaxies are classified as close-pair samples at $z = [4.5, 6.5]$, and 76 at $z = [6.5, 8.5]$.

\subsubsection{SFR enhancement as function of pair separation}
\label{sec: SFR enhancement small}
\begin{figure*}
    \centering
    \begin{subfigure}{0.48\linewidth}
        \centering
        \includegraphics[width=\linewidth, keepaspectratio]{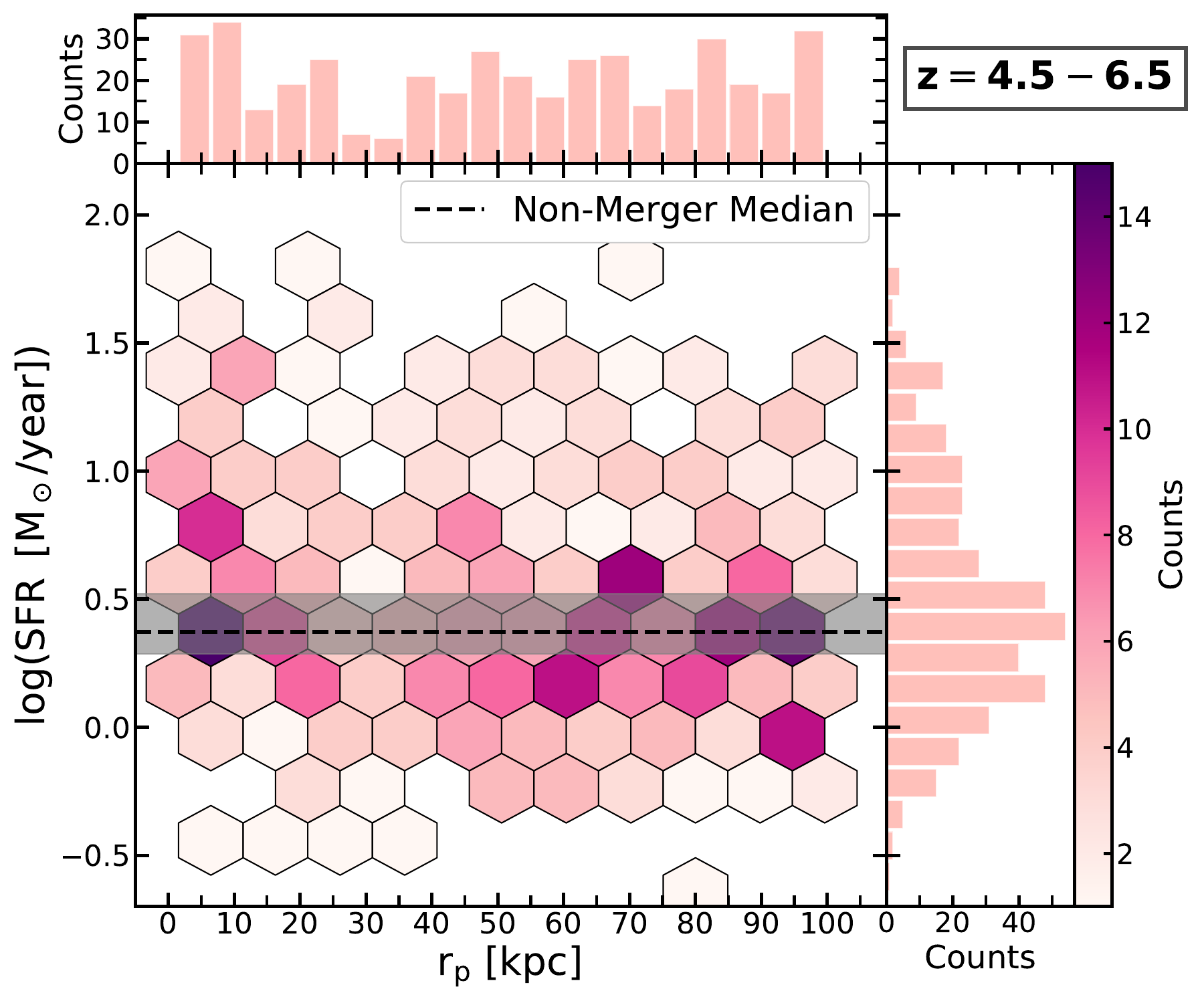} 
        \label{fig: SFR_hexbin_4.5_6.5}
    \end{subfigure}
    \hfill
    \begin{subfigure}{0.48\linewidth}
        \centering
        \includegraphics[width=\linewidth, keepaspectratio]{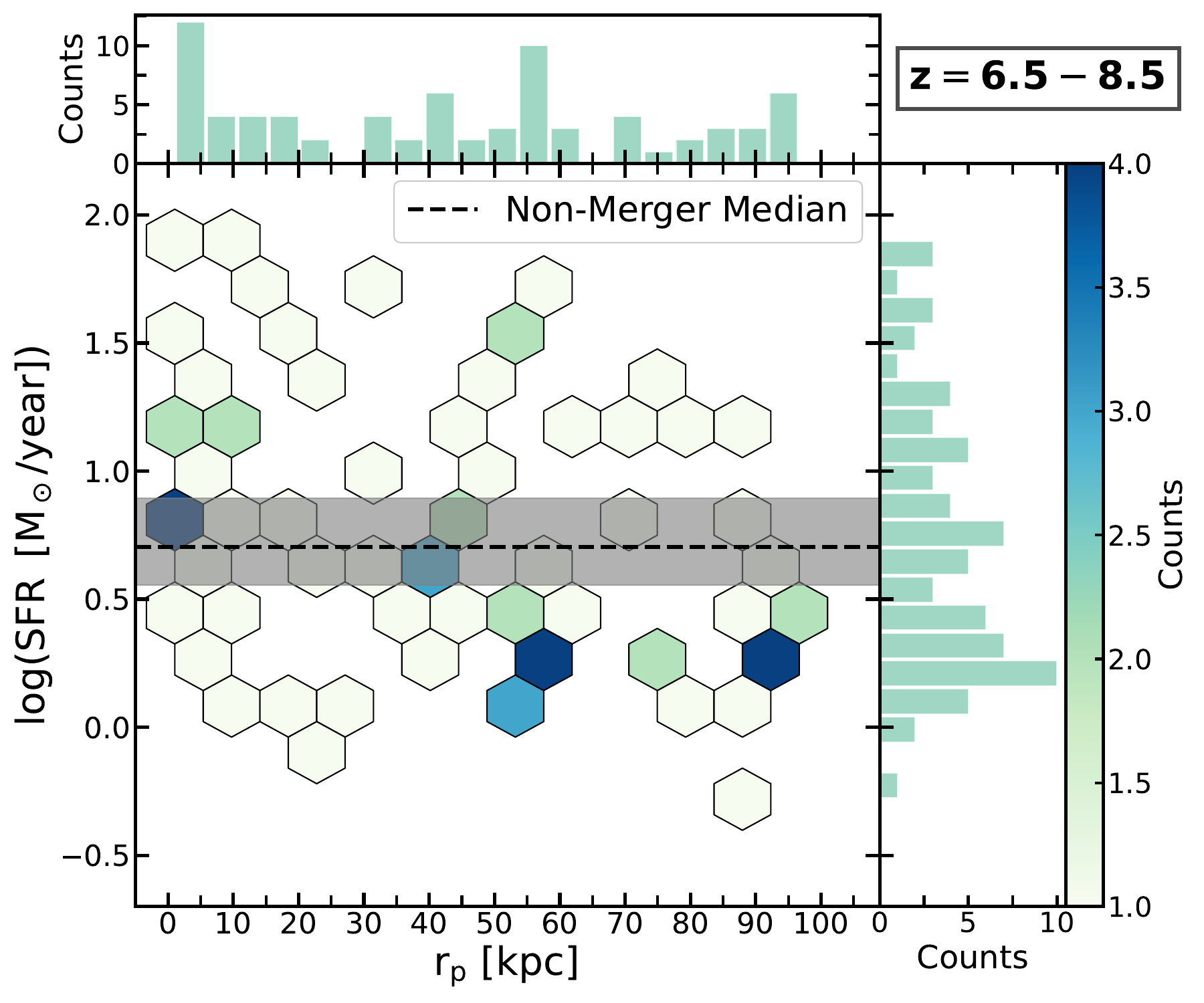} 
        \label{fig: SFR_hexbin_6.5_8.5}
    \end{subfigure}
    \caption{These two figures show the distribution of log(SFR) as a function of projected separation to the closest neighbor in close-pair samples at $z = [4.5, 6.5]$ (\textit{Left}) and $z = [6.5, 8.5]$ (\textit{Right}). Close-pair samples are defined as those with $r_p < 100$ kpc, $\Delta z < 0.3$, and a mass ratio of $\mu > 1/4$. In each hexbin, darker colors indicate a higher number of galaxies in that bin. The histograms at the top and right of each figure show the distributions of $r_p$ and log(SFR) for the close-pair samples, respectively.}
    \label{fig: SFR_Hexbin_all}
\end{figure*}

In this section, we present the results of our investigation into SFR enhancement. \autoref{fig: SFR_Hexbin_all} shows hexbin distributions of log(SFR) at various projected separations, up to 100 kpc, for all major merger close-pair samples across two redshift bins: $z = [4.5, 6.5]$ and $z = [6.5, 8.5]$. The median SFR of the non-merger samples is represented by a dashed line, with the 16th and 84th percentile ranges displayed as a shaded region. Since we do not have a statistically large enough dataset for every 10 kpc bin, we subsequently binned the data with a bin width of 20 kpc and used the bootstrapping method \citep{e89fac9c-03d7-3e22-aa30-08f5596f8fce,dca15e5b-b3f7-3417-8555-955fe36eb045} to calculate the SFR for each bin in order to maintain a sufficient number of sources in each bin. The results are illustrated in \autoref{fig: SFR vary redshift}. This figure also includes the SFRs of morphologically detected merging galaxies from the Gold Samples identified in \cite{dalmasso2024rate}, represented by gold symbols. The Gold Samples consist of galaxies whose morphology parameters satisfy the criteria from both \cite{2003AJ....126.1183C} and \cite{2008MNRAS.391.1137L} for classifying merging galaxies.

For $z = [4.5, 6.5]$, we detect enhancements in star formation rates only at $r_p < 20$ kpc bin, with an increase of $0.25 \pm 0.10$ dex above the median of the non-merger samples, corresponding to a factor of $1.77 \pm 0.35$. The SFR of close-pair samples at $r_p = 20 - 100$ kpc is comparable to the median value of non-merger samples but is below the 84th percentile of non-merger samples. Similarly, for $z = [6.5, 8.5]$, the enhancements in star formation rates are only seen at $r_p < 20$ kpc bin, with an increase of $0.26 \pm 0.11$ dex above the median of the non-merger samples, corresponding to a factor of $1.76 \pm 0.49$. No enhancement signals are detected at projected separations of $r_p = 20 - 100$ kpc. In particular, the SFR of close-pair samples at $r_p = [80, 100]$ kpc is approximately $0.36$ dex lower than the median value of non-merger samples. 

Since this is the first work to investigate SFR enhancement at the previously unexplored epoch of $z > 4.5$, we can only compare our findings with results at lower redshifts. \cite{2022ApJ...940....4S} used spectroscopic observations to investigate SFR enhancements at $0.5 < z < 3.0$, finding an enhancement factor of $\sim 1.23^{+0.08}_{-0.09}$ in the $r_p < 25$ kpc bin only. \cite{2014AJ....148..137L} identified merging galaxies in the COSMOS field between \(0.25 < z < 1.00\) with \(\log M_*/ \mathrm{M}_\odot > 10.6\), using an automated method to median-filter high-resolution HST images from COSMOS to distinguish two concentrated galaxy nuclei at small separations (2.2–8 kpc). They reported an SFR enhancement of \(2.1 \pm 0.6\) in their merging sample compared to their non-merging sample.  At low redshifts, the projected separation range for enhancement is generally larger. For example, \cite{ellison2008galaxy} found an enhancement range extending out to 30–40 kpc. \cite{2012MNRAS.426..549S} ($z < 1$) found statistically significant SFR enhancements extending to projected separations as far as 80 kpc. In contrast, \cite{2018ApJ...868...46S} found no significant star forming difference between merger and non-merger galaxies at $3 < r_p < 15$ kpc in the redshift range $0.3 < z < 2.5$, and \cite{2019ApJ...874...18W} did not observe significant SFR enhancements in merging galaxy pairs compared to control samples of isolated galaxies at \(1.5 \leq z \leq 3.5\).

High-redshift galaxies are generally more compact, with $R_e$ averaging around $1$ kpc, compared to $4$-$5$ kpc at lower redshifts ($z = 1$-$2$) \citep{2024MNRAS.527.6110O}. The smaller sizes of high-redshift galaxies naturally results in shorter permitted interaction ranges between close pairs. This size discrepancy may lead to a diminished scale of gravitational influence and less opportunity for gas inflow across extended distances. Furthermore, using SFRs calculated from other SFHs, including delayed, constant, and exponential models, we also observe enhanced SFRs in results derived from these alternative SFHs.

\subsubsection{sSFR and Stellar Mass Enhancement}

\begin{figure*}
    \centering
    \begin{subfigure}{\linewidth}
        \centering
        \includegraphics[width=\linewidth]{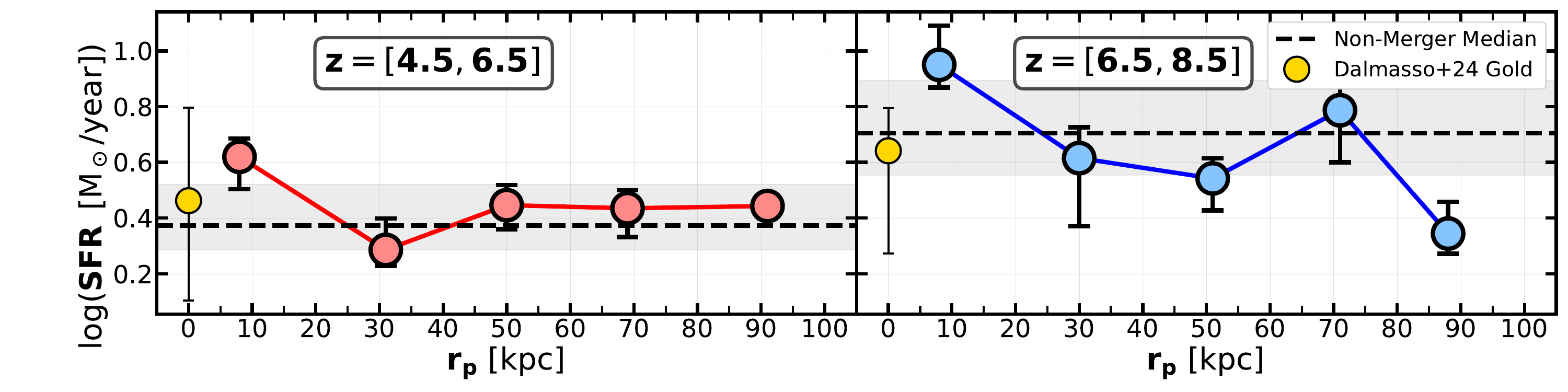}
        \caption{The average value of the star formation rate (in log units) as a function of projected separation to the closest neighbor in close-pair samples at $z = [4.5, 6.5]$ and $z = [6.5, 8.5]$. At $z = [4.5, 6.5]$, we observe an enhancement of $0.25 \pm 0.10$ dex above the non-merger median at $r_p < 20$ kpc. Similarly, at $z = [6.5, 8.5]$, an enhancement of $0.26 \pm 0.11$ dex is observed at $r_p < 20$ kpc. No enhancement signals are detected at other $r_p$ values in either redshift range.}
        \label{fig: SFR vary redshift}
    \end{subfigure}%
    \hfill
    \begin{subfigure}{\linewidth}
        \centering
        \includegraphics[width=\linewidth]{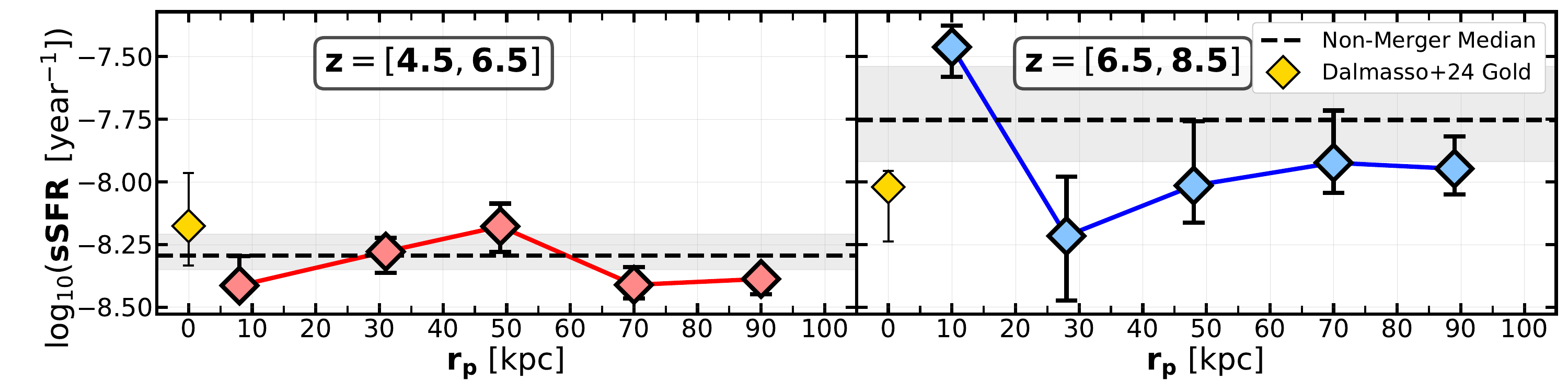}
        \caption{Average specific star formation rate (sSFR) as a function of projected separation to the closest neighbor in close-pair samples at $z = [4.5, 6.5]$ and $z = [6.5, 8.5]$. At $z = [4.5, 6.5]$, no enhancement in sSFR is observed across any projected separation range, with values oscillating around the non-merger median. At $z = [6.5, 8.5]$, an enhancement of $\sim 0.3$ dex above the non-merger median is detected at $r_p < 20$ kpc, while sSFR at $r_p > 20$ kpc shows values consistently below the non-merger median.}
        \label{fig: sSFR vary redshift}
    \end{subfigure}%
    \hfill
    \begin{subfigure}{\linewidth}
        \centering
        \includegraphics[width=\linewidth]{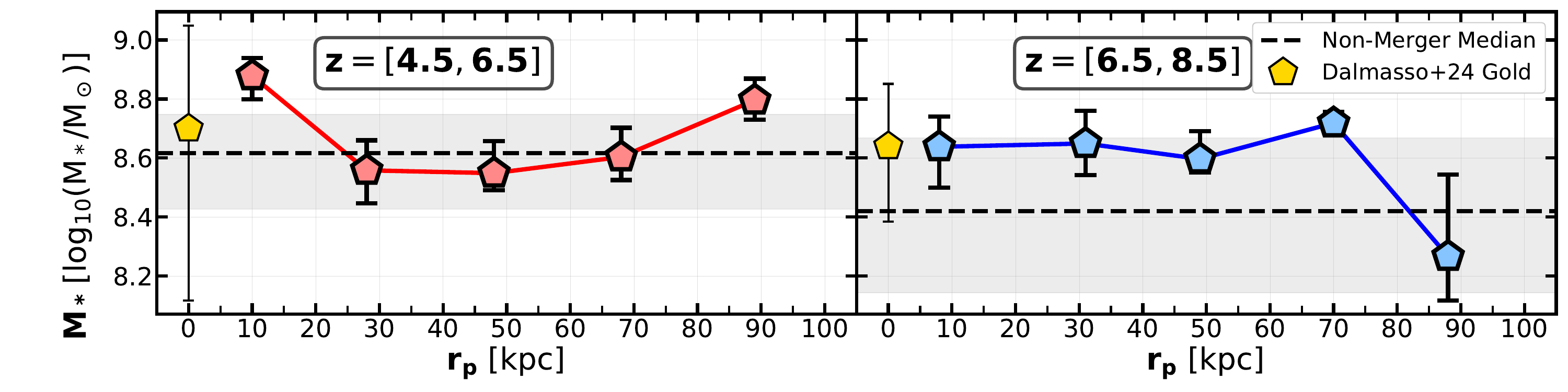}
        \caption{The average stellar mass for galaxies in pairs as a function of projected separation to the closest neighbor in close-pair samples at $z = [4.5, 6.5]$ and $z = [6.5, 8.5]$. At $z = [4.5, 6.5]$, an enhancement in stellar mass of 0.26 dex above the non-merger median is observed. At $z = [6.5, 8.5]$, the stellar mass of close-pair samples exceeds the non-merger median by 0.1 dex at $r_p < 80$ kpc. However, no strong evidence of significant stellar mass enhancement in close-pair galaxies compared to the non-merger sample is found.}
        \label{fig: stellar mass vary redshift}
    \end{subfigure}
    
    \caption{Three figures illustrate the evolution of galaxy SFR, sSFR, and stellar mass as a function of projected separation to the closest neighbor in close-pair samples computed from our observational data. The left panels present the results for $z = [4.5, 6.5]$, while the right panels show the results for $z = [6.5, 8.5]$. An enhancement in SFR is observed at $r_p < 20$ kpc. Results from morphology-identified merging samples from \protect\cite{dalmasso2024rate} are indicated by the yellow points at $r_p = 0$ kpc. Shaded regions represent the 16th and 84th percentile values.}
    \label{fig: combined vary redshift}
\end{figure*}

Having found some SFR enhancements at $r_p < 20$kpc bin in both redshift bins $z = [4.5, 6.5]$ and $z = [6.5, 8.5]$, we  investigate two additional intrinsic galaxy properties linked with star formation which may vary with $r_p$: specific star formation rate (sSFR) and stellar mass. These properties could provide further insights into star forming and stellar population variations at different projected separations and redshifts. The reason for investigating stellar mass enhancement is that typical merger timescales for close pairs at $z = 4.5 - 8.5$ last around $50 - 100$ Myr \citep{genel2014introducing, vogelsberger2014introducing, 2017MNRAS.468..207S, 2019ApJ...876..110D, 2022ApJ...940..168C, 2024arXiv240709472D}. It would be interesting to see whether the stellar mass increment from mergers during this period is significant or not. We apply the same method as in the SFR enhancement section (Section \ref{sec: SFR enhancement small}) to classify galaxies into close-pair and non-merger groups and investigate the variation of sSFR and stellar mass at different projected separations. The results are shown in \autoref{fig: sSFR vary redshift} and \autoref{fig: stellar mass vary redshift}, respectively.

\autoref{fig: sSFR vary redshift} shows the sSFR variations as a function of projected separation, $r_p$, for two redshift bins: $z = [4.5, 6.5]$ (left panel) and $z = [6.5, 8.5]$ (right panel). The red and blue diamonds represent the bootstrapped sSFR values for galaxies in close pairs across different bins, while the black dashed lines show the median sSFR for the non-merger sample. The shaded gray regions represent the 16th and 84th percentile ranges for the non-merger galaxies. Additionally, data points marked by gold diamonds correspond to the results from the \cite{dalmasso2024rate} Gold Samples, which identify morphologically detected merging galaxies. 

In the $z = [4.5, 6.5]$ bin, we observe that the sSFR values for close-pair galaxies slightly exceed the non-merger median in the $20 - 60$ kpc range, but remain within the 84th percentile range of the non-merger samples. No significant enhancement is detected at the smallest separation bin ($r_p < 20$ kpc). In the higher redshift bin, $z = [6.5, 8.5]$, the sSFR values show an enhancement at $r_p < 20$ kpc, with an increase of approximately $0.3$ dex relative to the non-merger median, while at $r_p > 20$ kpc, the sSFR values are consistently lower than those of the non-merger samples. Overall, neither redshift bin demonstrates a strong or consistent sSFR enhancement at any separation, and close-pair galaxies do not show a clear trend of significant sSFR elevation relative to the non-merger sample at $z = [4.5, 8.5]$.

In \autoref{fig: stellar mass vary redshift} we plot the variation of stellar mass as a function of projected separation, $r_p$, for two redshift bins: $z = [4.5, 6.5]$ (left panel) and $z = [6.5, 8.5]$ (right panel). In the $z = [4.5, 6.5]$ bin, we observe an enhancement in stellar mass at $r_p < 20$ kpc, with a value of approximately $0.26$ dex above the non-merger median. At $r_p = 20 - 80$ kpc, the stellar mass of close-pair galaxies is comparable to the median value of non-merger galaxies, while an additional enhancement of around $0.18$ dex is observed at $r_p = 80 - 100$ kpc. For the $z = [6.5, 8.5]$ bin, the stellar mass of close-pair galaxies exceeds the median value of the non-merger sample at $r_p < 80$ kpc, though it remains within the 84th percentile of the non-merger distribution. These results suggest that while stellar mass enhancement is observed at $r_p < 20$ kpc for $z = [4.5, 6.5]$ and at $r_p < 80$ kpc for $z = [6.5, 8.5]$, the enhancement is not statistically significant, and the range of projected separations where it occurs varies considerably. Therefore, we conclude that there is no strong evidence for significant stellar mass enhancement in close-pair galaxies compared to the non-merger sample.

The above SFR, sSFR, and stellar mass evolution with projected separation to the closest neighbor are all based on a fiducial line-of-sight difference of $\Delta z < 0.3$. In \autoref{fig: 3d 45 65} and \autoref{fig: 3d 65 85}, we present results using various $\Delta z$ upper limits, up to 0.5. Overall, we do not find noticeable differences between the results at different $\Delta z$ upper limits.

\subsubsection{Methodological Challenges in Comparing SFR Offset Between Close-Pair and Controlled Non-Merger Galaxy Samples}
\label{sec: Methodological Challenges in Comparing SFRs Between Merger and Controlled Non-Merger Galaxy Samples}
A more statistically robust method for investigating SFR enhancements involves comparing the SFR offsets between close-pair and contolled non-merger isolated samples, effectively minimizing biases related to stellar mass, redshift, and environmental factors \citep[e.g.,][]{perez2009building, 10.1111/j.1365-2966.2010.17076.x, 10.1111/j.1365-2966.2010.17932.x, 2012MNRAS.426..549S, 2021MNRAS.501.2969G, 2022ApJ...940....4S}. This approach requires constructing several controlled non-merger samples for each close-pair galaxy, matched closely in redshift and stellar mass. Low-redshift studies ($z < 1$) using SDSS data employ strict matching criteria when comparing galaxy samples. These studies \citep{10.1111/j.1365-2966.2010.17076.x, 10.1111/j.1365-2966.2010.17932.x, 2012MNRAS.426..549S} typically use a redshift difference limit of $\Delta z < 0.005$ and a stellar mass difference limit of $\Delta \mathrm{M}_* < 0.1$ dex to ensure close matching. For each iteration, one additional control sample is paired with each merger sample, and a Kolmogorov-Smirnov (K-S) test is performed on the distributions of stellar mass and redshift to ensure similarity. If the p-value remains high, the process is repeated until the similarity criterion is no longer met. At redshifts ($0.5 < z < 3$), \cite{2022ApJ...940....4S} used the criteria of a stellar mass difference within 0.15 dex, a spectroscopic redshift difference within 0.15, and an overdensity difference within a factor of 1 to create controlled isolated samples.

Given the limited size of our dataset at $z = [4.5, 6.5]$ (2404 galaxies) and $z = [6.5, 8.5]$ (926 galaxies), it is impractical to obtain a sufficient number of controlled non-merger isolated samples for each close-pair galaxy. We attempted this approach by applying broader matching criteria ($\Delta z < 0.1$ and $\Delta \mathrm{M}_* < 0.1$ dex), but this still resulted in a rapid decline in the p-value of the K-S test after only a few iterations. Specifically, the p-value for stellar mass dropped to 0.41 after seven iterations, and for redshift, it decreased to 0.25 after four iterations. Therefore, while this method is valuable, its effectiveness relies on having a sufficiently large sample size, which could be achieved with future deep and wide-field NIRCam observations, larger than those from CEERS, JADES, and NEP-TDF.

\subsubsection{Star Formation Trends from Semi-analytic Models}
\label{sec: star_formation_enhancement_simulation}

In this section, we investigate the evolution of SFR, sSFR, and stellar mass as a function of projected separation to the closest neighbor, using data from the \textit{Semi-analytic Forecasts for JWST} (SC-SAM) simulation. The \textit{Semi-analytic Forecasts for the Universe} project provides physically motivated predictions of galaxy and quasar properties and distributions across cosmic time, generated with the Santa Cruz semi-analytic model (SAM) for galaxy formation. This model strikes a balance between physical realism and computational efficiency and has been shown to successfully reproduce observational constraints up to redshift \(z \sim 10\) \citep{2019MNRAS.483.2983Y}.

The project delivers two main data products: \textit{Semi-analytic Forecasts for JWST} and \textit{Semi-analytic Forecasts for Roman}. In this study, we utilize the \textit{Semi-analytic Forecasts for JWST} dataset. The methodology and details of the simulation are well-documented across several studies \citep{2019MNRAS.483.2983Y,2019MNRAS.490.2855Y, 2020MNRAS.494.1002Y, 2020MNRAS.496.4574Y, 2021MNRAS.508.2706Y, 2022MNRAS.515.5416Y}

The \textit{deep} dataset comprises eight independent lightcone simulations. For each lightcone catalog, we apply a consistent mass completeness limit of $8 - 10$ [$\mathrm{log}_{10}(\mathrm{M}_* / \mathrm{M}_\odot)$], which initially filters the catalog and results in approximately 65,000 galaxies per lightcone. We then follow a methodology similar to that used for our observational data to classify galaxies into close-pair and non-merger samples. The only difference lies in the line-of-sight criterion: for observational data, we rely on $\Delta z$ due to the use of photometric redshifts, whereas in the simulation, exact redshift values and the large sample size allow us to apply a more stringent selection based on rest-frame relative velocity between galaxies, $\Delta v$, with a fiducial upper limit of 1000 km/s. 

\begin{figure*}
    \centering
    \begin{subfigure}{\linewidth}
        \centering
        \includegraphics[width=\linewidth]{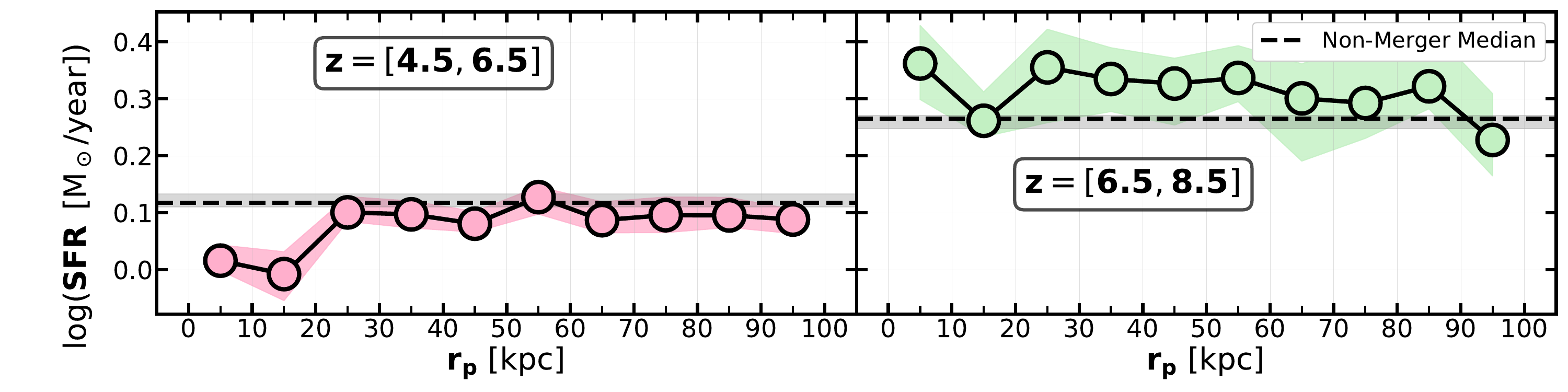}
        \caption{Averaged SFRs as a function of projected separation to the closest neighbor in close-pair samples at \( z = [4.5, 6.5] \) and \( z = [6.5, 8.5] \). No SFR enhancement signals are detected at any projected separation.}
        \label{fig: sc-sam SFR vary redshift}
    \end{subfigure}%
    \hfill
    \begin{subfigure}{\linewidth}
        \centering
        \includegraphics[width=\linewidth]{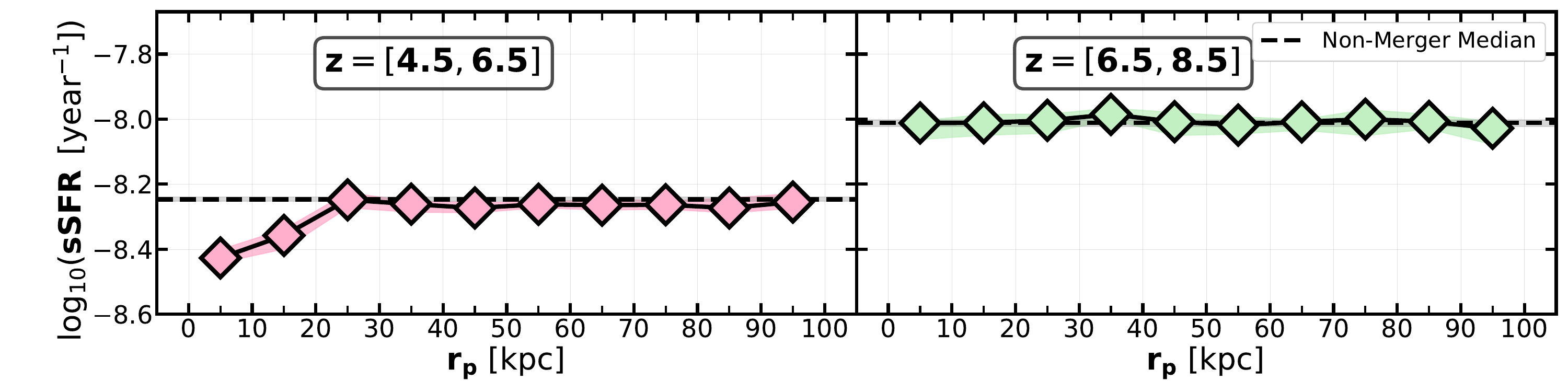}
        \caption{Averaged sSFRs as a function of projected separation to the closest neighbor in close-pair samples at \( z = [4.5, 6.5] \) and \( z = [6.5, 8.5] \). No enhancement signals are detected at any projected separation.}
        \label{fig: sc-sam sSFR vary redshift}
    \end{subfigure}%
    \hfill
    \begin{subfigure}{\linewidth}
        \centering
        \includegraphics[width=\linewidth]{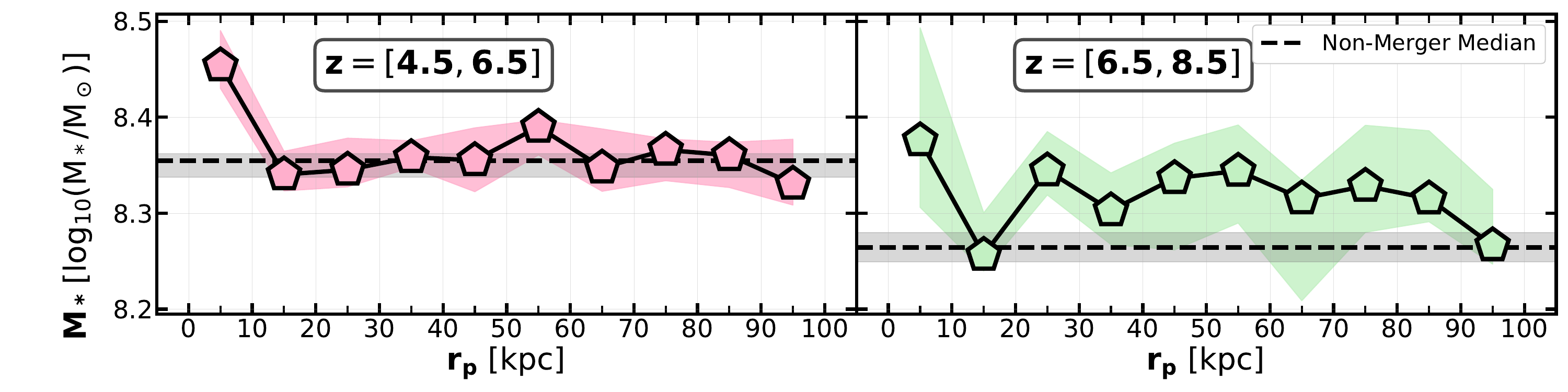}
        \caption{Average stellar mass as a function of projected separation to the closest neighbor in close-pair samples at \( z = [4.5, 6.5] \) and \( z = [6.5, 8.5] \). No strong evidence of significant stellar mass enhancement is observed.}
        \label{fig: sc-sam stellar mass vary redshift}
    \end{subfigure}
    
    \caption{Figures showing the evolution of galaxy SFR, sSFR, and stellar mass as a function of projected separation to the closest neighbor in close-pair samples with $\Delta v < 1000$ km/s computed from the SC-SAM simulation \textit{deep} dataset. The left panels presents the results for redshifts $z = [4.5, 6.5]$, while the right panels show the results for $z = [6.5, 8.5]$. We first compute these properties with projection separation for all eight realizations in the \textit{deep} catalog separately, using a bin width of 10 kpc and bootstrapping to obtain the median values. The results from all realizations are then combined, with the overall median calculated as the median of the eight realization medians. The 16th and 84th percentiles are derived from the distribution of the medians across all realizations, and are shown by the shaded regions. Overall, we do not find any star formation enhancements at any projected separation.}
    \label{fig: sc-sam combined vary redshift}
\end{figure*}

The variations in SFR, sSFR, and stellar mass as a function of projected separation to the closest neighbor, using a line-of-sight velocity difference of $\Delta v < 1000$ km/s, are shown in \autoref{fig: sc-sam SFR vary redshift}, \autoref{fig: sc-sam sSFR vary redshift}, and \autoref{fig: sc-sam stellar mass vary redshift}. We first compute these properties as a function of projection separation for all eight realizations separately, using a bin width of 10 kpc and bootstrapping to obtain the median values. The results from all realizations are then combined, with the overall median calculated as the median of the eight realization medians. Similarly, the 16th and 84th percentiles are derived from the distribution of the medians across all realizations. The left panels display the results for $z = [4.5, 6.5]$, while the right panels correspond to $z = [6.5, 8.5]$. Additionally, results using various $\Delta v$ upper limits, up to 5000 km/s, are shown in \autoref{fig: 3d 45 65 sc-sam} and \autoref{fig: 3d 65 85 sc-sam}. We also evaluated the results with $\Delta v$ upper limits of 100, 200, and 300 km/s for the SC-SAM data, and found them to be closely aligned with the results from other $\Delta v$ upper limits. As can be seen, we do not observe significant differences between the results across different $\Delta v$ upper limits.

At $z = [4.5, 6.5]$, we observe a flat trend in SFR from $r_p = 100$ kpc to 20 kpc, with values from the close-pair samples closely aligning with those of the non-merger samples. At $r_p < 20$ kpc, there is a slight decrease of 0.1 dex below the non-merger median. However, this trend does not hold at $z = [6.5, 8.5]$, where we find close-pair samples' SFRs are $0.06^{+0.05}_{-0.02}$ dex above the non-merger median across nearly all projection separation bins. Although there are some fluctuations in the close-pair samples relative to the non-merger medians, these fluctuations are small, and we do not find any significant star formation enhancement at any projected separation and redshift bin.

For stellar mass at $z = [4.5, 6.5]$, we observe an opposite trend to the SFR variations in this redshift range. The stellar mass oscillates around the non-merger median from $r_p = 100$ kpc to $r_p = 10$ kpc, with amplitude of approximately 0.05 dex. At $r_p < 10$ kpc, an increase of 0.1 dex in stellar mass is observed. The stellar mass evolution at $z = [6.5, 8.5]$ follows a similar trend to the SFR evolution in this redshift range.

Since sSFR is defined as the ratio of SFR to stellar mass, the observed trends at $z = [4.5, 6.5]$—a decrease in SFR at $r_p < 20$ kpc and an increase in stellar mass at $r_p > 10$ kpc—result in a decrease in sSFR at $r_p < 20$ kpc. Between $r_p = 20$ and 100 kpc, the sSFR of the close-pair samples closely aligns with the non-merger medians. At $z = [6.5, 8.5]$, sSFR remains close to the non-merger medians at all projection separations.

Overall, while we observe some variations in SFR, sSFR, and stellar mass in the close-pair samples relative to the non-merger samples at $r_p < 20$ kpc from SC-SAM simulation, these variations are not significant and do not indicate any star formation enhancement at any projected separation. This suggests that the gas inflow resulting from mergers may be underestimated in the simulation, and thus does not sufficiently trigger star formation.

\subsection{Mergers Trigger AGN Activity}
\label{sec: mergers triggered agn activity}
The unexpectedly high population of AGNs and gravitational wave detections in the early universe remains a mystery. High-redshift galaxies alone are not expected to form AGNs within such short timescales at these early epochs. One possible explanation is the frequent occurrence of galaxy mergers. Galaxy merger rates at $z < 6$ have been extensively investigated \citep[e.g.,][]{2002ApJ...565..208P, 2003ApJS..147....1C, 2007ApJ...659..931B, conselice2008structures, 2015A&A...576A..53L, 2016ApJ...830...89M, 2017MNRAS.470.3507M, 2017A&A...608A...9V, 2018MNRAS.475.1549M, 2019ApJ...876..110D, 2022ApJ...940..168C}, though the galaxy merger rate at very high redshift ($z > 6$) had not been studied in detail until the recent observational study by \cite{2024arXiv240709472D} and \cite{2025arXiv250201721P}. They presented pair fractions, merger rates, and stellar mass accretion rates at $z > 6$ revealing rapid galaxy merger rates of approximately 6 mergers per Gyr. In this section, building on their results, we further investigate the AGN fractions and AGN excess in close-pair samples.  As in addition to star formation rate, it is possible that close pairs of galaxies and mergers will induce AGN activity in galaxies. 

\subsubsection{AGN Identification}
In addition to the photometric data, we incorporate public NIRSpec spectroscopic data to identify AGNs. Unlike the previous sections on merger-triggered star formation properties, where we utilized data from all eight fields (CEERS, JADES GOODS-S, NEP-TDF, NGDEEP, GLASS, El-Gordo, SMACS-0723, MACS-0416), only the CEERS and JADES GOODS-S fields contain a substantial number of spectroscopic observations. Therefore, this section focuses on AGN fractions and AGN excess in these two fields. We employ two independent methods to identify AGNs: the photometric SED method and the spectroscopic BPT diagnostic method. The photometric SED method, along with the relevant AGN templates, is outlined in \cite{2023MNRAS.525.1353J}. We will discuss the BPT method in detail and present results from both methods. At high redshift ($z > 3$), the rapid evolution towards lower metallicities and higher ionization parameters introduces challenges in identifying AGNs using the BPT diagram. However, some studies have demonstrated that AGN detection through BPT diagnostics remains possible even in these conditions \citep[e.g.,][]{2023A&A...679A..89P, 2023arXiv231118731S, 2024arXiv240815615M}.

Out of the 2525 PRISM spectra in the JADES GOODS-S catalog, we use all spectra with spectroscopic redshifts categorized as A, B, or C. These categories are defined by the presence of at least two emission lines or a combination of a line and a strong continuum break in the PRISM spectra \citep{bunker2023jades, 2024arXiv240406531D}. After applying this criterion, 1333 galaxies remain. We then apply the BPT diagnostic diagram \citep{1987ApJS...63..295V, 2001ApJ...556..121K, 2003MNRAS.346.1055K, 2006MNRAS.372..961K} to identify AGN based on line ratios of H$\beta$ \(\lambda4861\) / [O~\textsc{iii}] \(\lambda5007\) and [N~\textsc{ii}] \(\lambda6583\) / H$\alpha$. The resolution of PRISM spectra ($R \sim 100$) is not high enough to separate [N~\textsc{ii}] \(\lambda6583\) from H$\alpha$. Therefore, we calculate the [N~\textsc{ii}] \(\lambda6583\) line flux by estimating the H$\alpha$ line flux from the H$\beta$ flux, assuming a dust-free line ratio of 2.86, and subtracting this value from the combined flux of H$\alpha$ and [N~\textsc{ii}] \(\lambda6583\).

Given the wavelength coverage of NIRSpec PRISM spectra (6000 \, \text{\AA} to 53,000 \, \text{\AA}), we can simultaneously detect the aforementioned AGN diagnostic lines only within the redshift range of \(5.3 < z < 6.7\). In total, we identify 40 AGNs in GOODS-S, of which 19 are successfully matched with the EPOCHS photometric catalog.

\begin{figure}
    \centering
    \includegraphics[width=\linewidth]{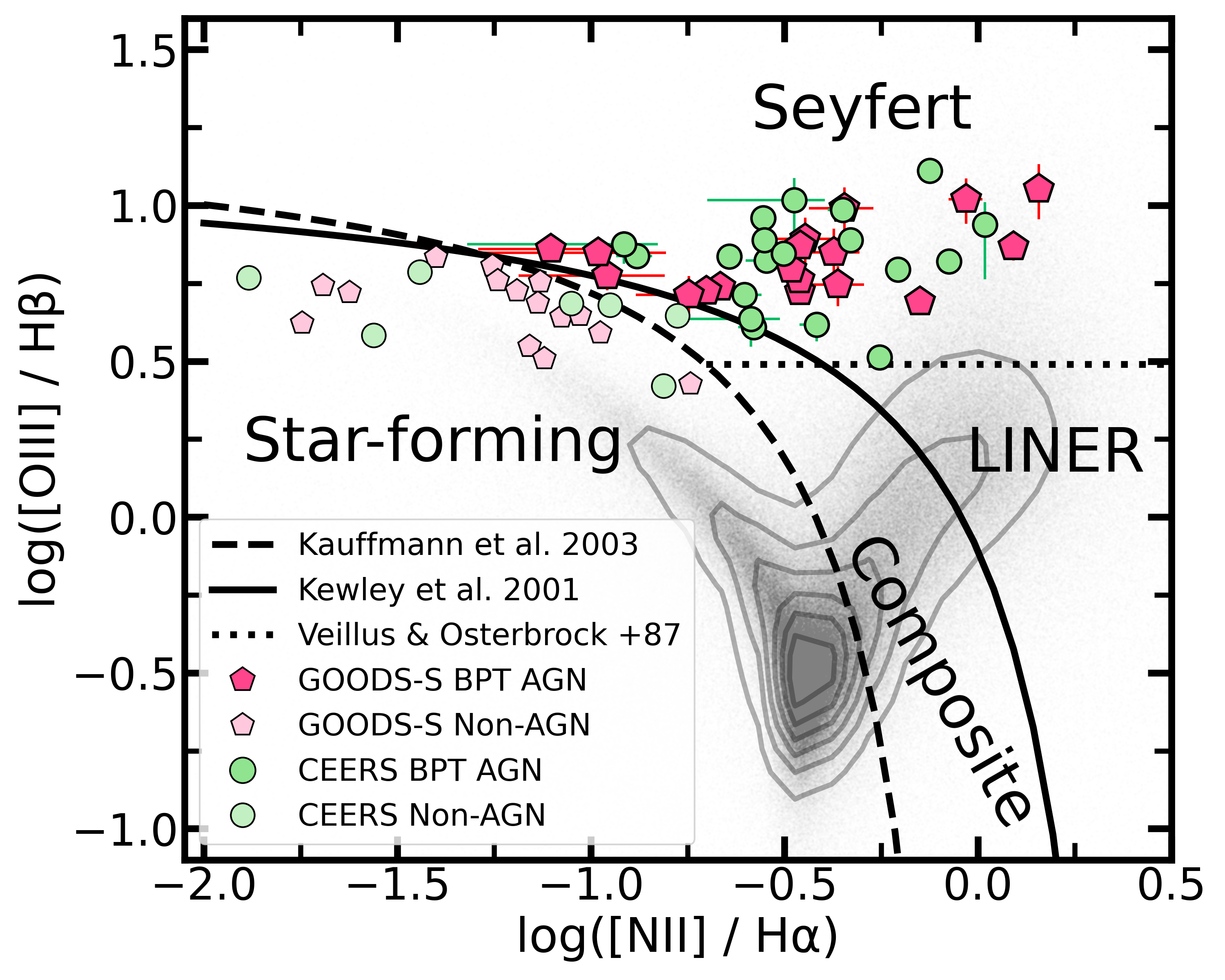}
    \caption{BPT diagnostic diagram, based on criteria from \citet{1987ApJS...63..295V, 2001ApJ...556..121K, 2003MNRAS.346.1055K, 2006MNRAS.372..961K}. Galaxies classified as AGNs are shown as red pentagons for the JADES GOODS-S field and green circles for the CEERS field, while non-AGNs are labeled with pink pentagons for JADES GOODS-S and light-green circles for CEERS. The contours are derived from 927,552 low-redshift SDSS DR7 galaxies, as documented by \citet{2009ApJS..182..543A}.}
    \label{fig: BPT Nii Halpha}
\end{figure}

Additionally, by cross-matching the EPOCHS photometric catalog with BPT-confirmed AGNs from the CEERS PRISM spectra catalog produced by DAWN, we identify and match 25 AGNs. The positions of these 44 AGNs in the BPT diagram are marked with pink pentagons (for JADES) and green circles (for CEERS) in \autoref{fig: BPT Nii Halpha}.

\subsubsection{AGN Fraction of Galaxies in Pairs}

We follow criteria (i) and (ii) outlined in \autoref{sec: close-pair classifications} to classify galaxies into close-pair and non-merger samples. Close-pair galaxies are defined as those with $\Delta z < 0.3$ and $r_p < 100$ kpc. The reason for not applying criterion (iii) — the major merger criterion — is that our goal is to explore all possible AGN presence in pairs, regardless of the merger mass ratio. We also do not have a large enough sample to investigate differences in mass ratios and separations.  We thus are searching for a detection of significance, rather than one of measurement given our limited data.

We present three different AGN fractions in this work. When we refer to AGN we are only including the measurements of spectra and imaging that imply these systems are AGN as we do not have a complete sample of these systems. The quantities we measure are defined as follows:
\begin{itemize}
    \item \( f_{\mathrm{AGN, AGN}} \): the fraction of AGNs in pairs relative to the total number of AGNs, calculated by dividing the number of AGNs in pairs by the total number of AGNs.  For example, if this value is $= 1$ this would imply that all AGN are in a galaxy pair, likewise if $=0$ no AGN are found in any galaxies in pairs. 
    \item \( f_{\mathrm{AGN, close-pair}} \): the fraction of AGNs in close-pair merger samples, calculated by dividing the number of AGNs in pairs by the total number of galaxies in pairs.  A value $= 1$ would mean that all galaxies in pairs have an AGN, and $=0$ would imply that no galaxies in pairs contain an AGN.
    \item \( f_{\mathrm{AGN, isolated}} \): the fraction of AGNs in non-merger samples, calculated by dividing the number of AGNs not in pairs by the total number of galaxies not in pairs.  Similar to above, a value $=1$ would imply all isolated galaxies have an AGN and $=0$ would imply no isolated galaxies have an AGN.
\end{itemize}

\begin{figure*}
    \centering
    \begin{subfigure}{\linewidth}
        \centering
        \includegraphics[width=\linewidth, height=0.23\paperheight, keepaspectratio]{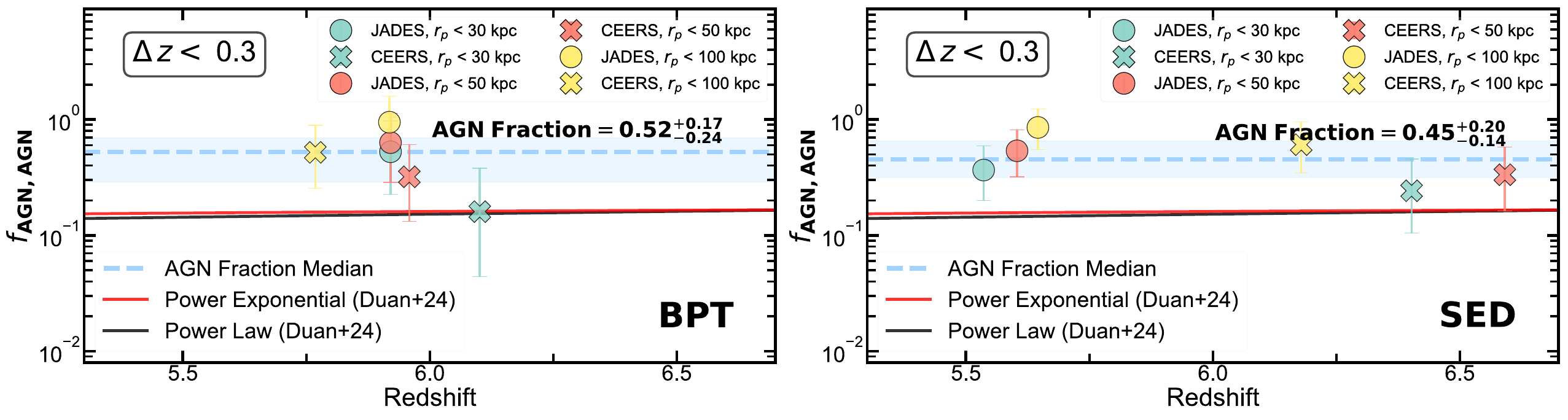} 
        \caption{Fraction of AGNs in pairs relative to the total number of AGNs, calculated by dividing the number of AGNs in pairs by the total number of AGNs. Circles represent JADES AGNs and crosses denote CEERS AGNs. AGN fractions calculated with projected separation upper limits of 30, 50, and 100 kpc are colored in green, red, and yellow, respectively. We find that the fraction of AGNs in pairs is $3.25^{+1.50}_{-1.06}$ times higher than the fraction of galaxies in pairs, with almost all AGNs having a companion within 100 kpc.}
        \label{fig: AGN_Fraction_In_pair}
    \end{subfigure}

    \begin{subfigure}{\linewidth}
        \centering
        \includegraphics[width=\linewidth, height=0.23\paperheight, keepaspectratio]{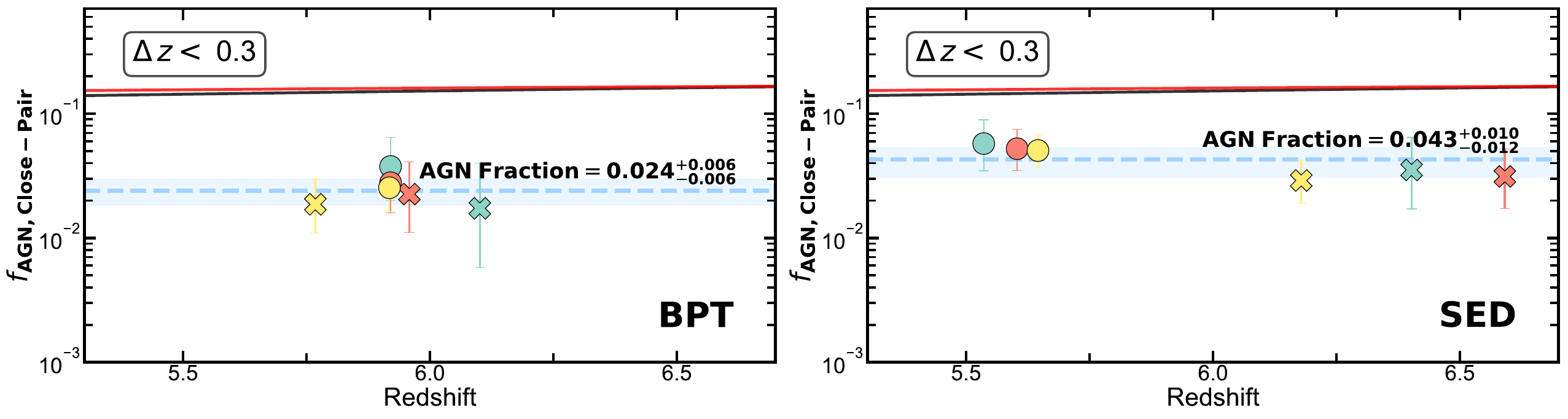} 
        \caption{The fraction of AGNs in close-pair merger samples, calculated by dividing the number of AGNs in pairs by the total number of galaxies in pairs. Circles indicate JADES AGNs, while crosses represent CEERS AGNs. AGN fractions calculated with projected separation upper limits of 30, 50, and 100 kpc are colored in green, red, and yellow, respectively.}
        \label{fig: AGN_pair_galaxy_pair}
    \end{subfigure}

    \begin{subfigure}{\linewidth}
        \centering
        \includegraphics[width=\linewidth, height=0.23\paperheight, keepaspectratio]{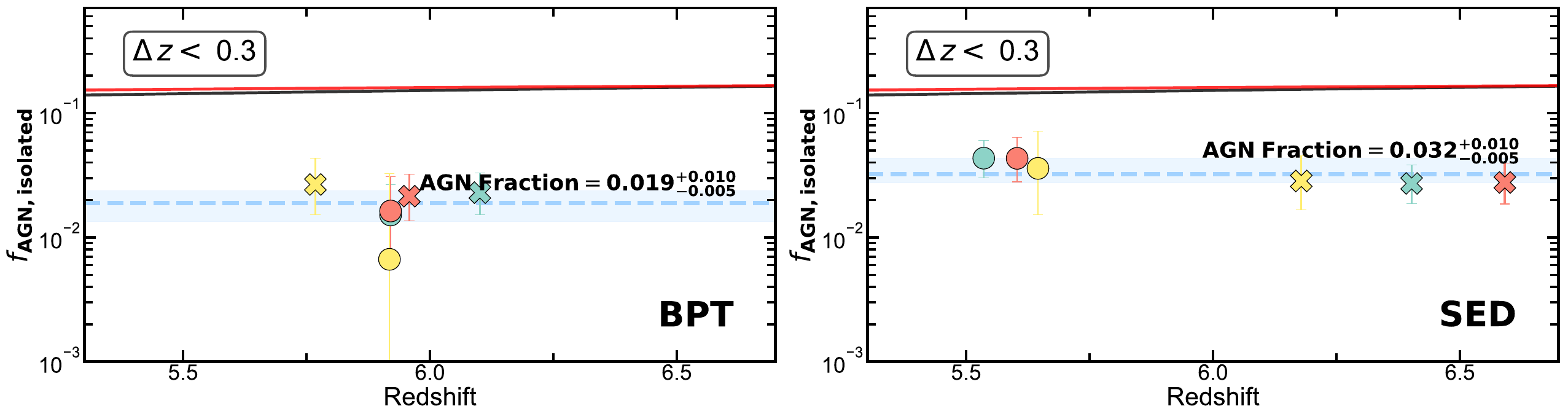}
        \caption{The fraction of AGNs in non-merger samples, calculated by dividing the number of AGNs not in pairs by the total number of galaxies not in pairs. Circles indicate pairs from the JADES data set and crosses represent pairs from the CEERS data set. AGN fractions calculated with projected separation upper limits of 30, 50, and 100 kpc are colored in green, red, and yellow, respectively.}

        \label{fig: AGN_agn_isolated_galaxy_isolated}
    \end{subfigure}
    \caption{Three figures showing three different AGN fractions: \( f_{\mathrm{AGN, AGN}} \), \( f_{\mathrm{AGN, close-pair}} \), and \( f_{\mathrm{AGN, isolated}} \) in our samples. Results for AGNs identified using the BPT diagnostic diagram are presented in the left figures, while results for AGNs selected using photometric SED are shown in the right figure. JADES GOODS-S data points are shown as circles, and CEERS points are shown as crosses. For each figure, the median AGN fractions are indicated by blue dotted lines, with shaded regions representing the 16th and 84th percentile values. Galaxy pair fractions from the power-law model and the power-law + exponential model from \protect\citep{2024arXiv240709472D} are shown as black and red lines. An AGN excess of $1.26 \pm 0.06$ and $1.34 \pm 0.06$ is detected using the BPT and SED methods, respectively. We do not observe an increase in AGN fraction as close-pair separation decreases.}
    \label{fig:combined}
\end{figure*}

We present these fractions \( f_{\mathrm{AGN, AGN}} \), \( f_{\mathrm{AGN, close-pair}} \), and \( f_{\mathrm{AGN, isolated}} \), alongside the redshift evolution of galaxy pair fractions modeled using both the power-law and power-law + exponential models from \cite{2024arXiv240709472D}, in \autoref{fig: AGN_Fraction_In_pair}, \autoref{fig: AGN_pair_galaxy_pair}, and \autoref{fig: AGN_agn_isolated_galaxy_isolated}. The left panel of each figure shows the AGN fraction for AGNs identified using the BPT diagram, while the right panel shows the AGN fraction for AGNs identified using the photometric SED method. AGN fractions from JADES GOODS-S are represented by circles, while those from CEERS are indicated by crosses. The colors correspond to the maximum projected separation between pairs: green, red, and yellow represent separations of 30, 50, and 100 kpc, respectively. The median value of all data points in each AGN fraction figure is represented by a blue dotted line, with the 16th and 84th percentile values serving as the $1\sigma$ confidence intervals. Given the small number counts where distributions do not follow a perfect Poisson distribution, errors for the AGN fractions are computed using the method described by \cite{1986ApJ...303..336G}. 

A direct result that can be drawn from \autoref{fig: AGN_Fraction_In_pair} is that half of the AGNs are found in pairs at \( r_p < 50 \, \text{kpc} \) and \( \Delta z < 0.3 \). This fraction (\( f_{\mathrm{AGN, AGN}} \)) is $3.25^{+1.50}_{-1.06}$ times higher the galaxy pair fractions, even at the smallest projection separation limit of $r_p < 30$ kpc. The distribution of $r_p < 100$ kpc points shows that almost all AGNs have a companion within 100 kpc in projection and within $\Delta z < 0.3$ along the line-of-sight. Given the high merger rate in this redshift range ($\sim 6$ per Gyr) from \cite{2024arXiv240709472D}, this suggests that merger-triggered AGN activity is a consistent process. After two galaxies have merged and triggered an AGN, it is highly likely that there is another galaxy within 100 kpc that will eventually merge with the product of the previous merger, further triggering AGN activity. 

Furthermore, the fraction of AGNs in all close-pair samples, \( f_{\mathrm{AGN, close-pair}} \), shows a modest difference between the BPT and SED methods. The BPT method gives a value of $0.024^{+0.006}_{-0.006}$, while the SED method yields $0.043^{+0.010}_{-0.012}$. This slight discrepancy is expected, as the AGNs identified by the BPT and SED methods have an unknown overlap which is certainly not identical. This is also reflected in \autoref{fig: AGN_agn_isolated_galaxy_isolated}, where we present \( f_{\mathrm{AGN, isolated}} \), the fraction of AGNs in isolated samples. For isolated galaxies, the BPT method gives $0.019^{+0.005}_{-0.005}$, while the SED method finds $0.032^{+0.011}_{-0.005}$.

Previous studies at low redshift using SDSS data, such as \cite{ellison2011galaxy, satyapal2014galaxy}, show that \( f_{\mathrm{AGN, close-pair}} \) increases with decreasing $r_p$, rising by a factor of up to $\sim 2.5$ at $r_p < 10$ kpc. However, we do not observe such a trend in our data due to the limited sample size, which hinders our ability to trace this projected separation evolution in detail.

Despite the differences in AGN fractions between the two methods, the key parameter in this section is the AGN excess, defined as the ratio \( f_{\mathrm{AGN, close-pair}} / f_{\mathrm{AGN, isolated}} \).   We call this the ``AGN Excess'' which is in many ways a better quantity to determine than the individual AGN fractions. The reason for this is that the methods for locating and identify AGN is bound to have errors, biases, incompleteness and contamination. To first order we can assume that these biases are the same for those in pairs as compared to those which are isolated. The AGN excess from the BPT method is $1.26 \pm 0.06$, while the SED method shows a value of $1.34 \pm 0.06$. Although the AGN excess values are not highly significant and come with considerable uncertainties, the consistent detection of AGN excess from two independent methods supports the robustness of this result of a slight, but significant excess. Our result 

Compared to other studies on AGN excess, the work by \cite{2020A&A...637A..94G}, which utilized the SDSS and GAMA surveys to study AGN fractions up to \(z = 0.6\), found an AGN excess up to $1.81 \pm 0.16$. Using MaNGA data,  \cite{2024ApJ...963...53C} found an AGN excess of $\sim 1.8$ in major mergers and 1.5 in minor mergers. \cite{2024A&A...684A.111K} concluded that the excess of AGN activity in cluster outskirts is likely due to mergers and disturbed hosts, based on a sample of 19 thoroughly studied X-ray-selected galaxy clusters from the LoCuSS survey. All these clusters are considered massive, with $M_{500} \gtrsim 2 \times 10^{14} M_\odot$ at $z = 0.16 - 0.28$.

These studies used large sample sizes to compute the AGN excess. In our case, given the limitations in the number of samples, the AGN excess found in our work is likely a lower limit, with the actual AGN excess potentially similar or higher than those reported in aforementioned studies at low-redshift.

\section{Discussion}
\label{sec: Discussion}

While this work represents an early attempt to investigate the impact of mergers on galaxy star formation and AGN properties at $z > 4.5$, several aspects and limitations should be carefully considered.

First, this study is based on photometric redshifts. As a result, we adopt a redshift difference for the line-of-sight criterion rather than a stricter velocity difference. Without precise spectroscopic redshifts, some of the close-pair systems identified may not actually be physically close and may have significant redshift differences. Although we utilized an extensive datasets available to date (CEERS, JADES GOODS-S, NEP-TDF, NGDEEP, GLASS, El-Gordo, SMACS-0723, and MACS-0416), the sample size is still limited, and the lack of spectroscopic redshifts is a more prominent drawback. Based on our observationl results on SFR, sSFR, and stellar mass variations (\autoref{fig: SFR vary redshift}, \autoref{fig: sSFR vary redshift}, \autoref{fig: stellar mass vary redshift}), apart from the SFR enhancement at $r_p < 20$ kpc observed in both redshift bins $z = [4.5, 6.5]$ and $z = [6.5, 8.5]$, the other results do not show statistically significant evolutionary trends.

Second, the SFR, sSFR, and stellar mass values in this work are derived from photometric SED fitting, which has been shown to have a correlation coefficient of approximately 0.60 with SED fitting using spectroscopic data \citep{2024MNRAS.529.4728D}. While this moderate correlation does not heavily impact our results, the lack of a large number of spectroscopic observations limits our ability to refine these measurements. If a substantial amount of spectroscopic data were available, we could compute the SFR using Balmer lines and reanalyze the data to verify whether the observed SFR enhancement at $r_p < 20$ kpc remains consistent, whether the enhancement occurred at larger projection separation values, and whether close-pair samples present a bursty star formation compared to the non-merging sample.

An interesting finding in our study is that the SFR of morphologically identified merging galaxies is lower than that of close-pair galaxies at $r_p < 20$ kpc. This contrasts with previous results. \cite{2022ApJ...940....4S} found that morphologically identified merging galaxies consistently exhibited higher star formation enhancement than close-pair samples at $0.5 < z < 3.0$. There are two potential explanations for this discrepancy. First, due to the faint and distant nature of high-redshift galaxies, even with JWST's resolution, the morphological parameters used in \cite{dalmasso2024rate}, such as the CAS and Gini coefficients, suffer from high uncertainties, which could lead to falsely identifying merging galaxies that are, in fact, single galaxies. Second, it is possible that the majority of morphology-selected galaxies are in the very late stages after coalescence \citep{ferreira_2024}. \cite{ellison2022galaxy} showed that after the merger process is complete, mergers can rapidly quench star formation.

In addition to the observational data, we examine the evolution of star-forming properties using the SC-SAM simulation. In the simulation, we do not detect any significant star formation enhancement at any projected separation from the closest pair. The overall evolution of SFR, sSFR, and stellar mass with projected separation exhibits a largely flat trend, with only minor variations in the smallest bin, $r_p < 20$ kpc, relative to the non-merger median values. However, these deviations are not statistically significant. While the observational stellar mass evolution shows modest alignment with the simulation results, the SFR and sSFR differ considerably. This suggests that the influence of mergers on star-forming properties, particularly processes such as gas inflow, may be underestimated in the SC-SAM simulation.

Lastly, we investigate AGN Fraction and AGN Excess at $z = [5.3, 6.7]$ for the first time. We detect an AGN Excess of $1.26 \pm 0.06$ and $1.34 \pm 0.06$ in close-pair samples relative to non-merger samples for AGNs selected using the BPT diagnostic diagram and SEDs, respectively. Although the excess is not highly significant due to the limited size of our dataset, the consistent detection of AGN Excess from two independent methods strengthens the robustness of these results. One could argue that AGNs selected from BPT and photometric SED methods are less persuasive than broad-line AGNs (BL AGNs). However, detecting BL AGNs is another significant task and beyond the scope of this paper. To date, no comprehensive BL AGN catalogs have been published at these redshifts. Nevertheless, the AGN Excess detected from both the photometric SED and BPT methods lends solid support to the idea that at high redshift AGNs are triggered in part by the merging process.

Overall, our results are not significantly different in terms of overall findings compared to lower redshift galaxies \citep[e.g.,][]{ellison2008galaxy,2012MNRAS.426..549S, satyapal2014galaxy, 2022ApJ...940....4S, 2024ApJ...963...53C}.  We find an excess of star formation, that is not quite significant except for the closest galaxies ($r_p < 20$ kpc) at the smallest differences in redshift, between pairs. We also find a significant excess of AGNs in pairs compared to those in the field.  This excess is only at most a few 10s\%, thus less than a factor of two. It remains to be determined how quantitatively this excess results in the growth in super massive black holes.  In the future when better data and spectroscopic redshifts are available at these higher redshifts we can determine with more certainty the role of mergers and galaxies in pairs in producing star formation and AGN activity.

\section{Conclusions}
\label{sec: Conclusions}

In this paper, we present an early analysis of the impact of mergers on galaxy star-forming properties and AGN activity at $z = 4.5 - 8.5$. We explore in detail how galaxy SFR, sSFR, and stellar mass evolves as a function of projected separation to the closest neighbor, and whether these properties show any enhancement compared to non-merger samples in both observational and simulation data. Additionally, we examine how the AGN fraction varies between close-pair and non-merger samples and assess the AGN excess in these fractions. Our key findings are summarized as follows:

I. We observe an enhancement in SFR from the observational data at $r_p < 20$ kpc. The extent of this enhancement is $0.25 \pm 0.10$ dex above the non-merger median for $z = [4.5, 6.5]$ and $0.26 \pm 0.11$ dex for $z = [6.5, 8.5]$. No enhancement signals are detected at larger separations.  We thus conclude that there is a marginal excess in star formation for very close pairs of galaxies. 

II. Comparing this work's results with simulations we show that star formation properties derived from the SC-SAM simulation show no significant evolution or variation at $r_p < 100$ kpc. That is, no SFR enhancement signal is detected at $r_p < 20$ kpc in either redshift bin, $z = [4.5, 6.5]$ or $z = [6.5, 8.5]$.

III. The sSFR values in the simulations shows minimal variation at all projected separations and redshift ranges. Stellar mass from the simulations shows a small enhancement of approximately 0.1 dex above the non-merger median at $r_p < 20$ kpc.

IV. We compute the fraction of AGNs in pairs relative to the total number of AGNs (\( f_{\mathrm{AGN, AGN}} \)) and find that this fraction is $3.25^{+1.50}_{-1.06}$ times higher than the galaxy pair fraction at the same redshift. Additionally, almost all AGNs have a companion within 100 kpc.

V. We investigate AGN excess in close-pair samples relative to non-merger samples and find an excess of $1.26 \pm 0.06$ for BPT-selected AGNs and $1.34 \pm 0.06$ for photometric SED-selected AGNs. While the excess is not highly significant due to the limited size of our dataset, the consistent detection of AGN excess from two independent methods supports the robustness of these results.

Overall, using currently one of the most extensive dataset from eight JWST observational fields (CEERS, JADES GOODS-S, NEP-TDF, NGDEEP, GLASS, El-Gordo, SMACS-0723, and MACS-0416), we explore merger-drive galaxy star forming and AGN activities at $z = [4.5, 8.5]$ for the first time. This study currently represents one of the highest redshift analysis of mergers and may continue to be one of the most advanced. 

A worthwhile future endeavor, when a statistically substantial number of JWST spectroscopic observations are available, is to investigate this formation and activity using spectroscopic data e.g., balmer lines for SFR calculations and AGN identification.   We are also strongly limited to relatively lower mass galaxies within this study.  Ideally we would also like to probe the more massive galaxies, but these are very rare at these redshifts.  This will require larger surveys with JWST or in the future with Euclid/Roman. Likewise, deeper JWST surveys are needed to obtain complete samples of galaxies, as at the current time we are incomplete for most masses at the highest redshifts where our distant galaxy sample is measured.

\section*{Acknowledgements}

We acknowledge support from the ERC Advanced Investigator Grant EPOCHS (788113), as well as a studentship from STFC. LF acknowledges financial support from Coordenação de Aperfeiçoamento de Pessoal de Nível Superior - Brazil (CAPES) in the form of a PhD studentship. CCL acknowledges support from the Royal Society under grant RGF/EA/181016. CT acknowledges funding from the Science and Technology Facilities Council (STFC). This work is based on observations made with the NASA/ESA \textit{Hubble Space Telescope} (HST) and NASA/ESA/CSA \textit{James Webb Space Telescope} (JWST) obtained from the \texttt{Mikulski Archive for Space Telescopes} (\texttt{MAST}) at the \textit{Space Telescope Science Institute} (STScI), which is operated by the Association of Universities for Research in Astronomy, Inc., under NASA contract NAS 5-03127 for JWST, and NAS 5–26555 for HST. Some of the data products presented herein were retrieved from the Dawn JWST Archive (DJA). DJA is an initiative of the Cosmic Dawn Center, which is funded by the Danish National Research Foundation under grant No. 140. This research made use of the following Python libraries: \textsc{Numpy} \citep{harris2020array}; 
\textsc{Scipy} \citep{2020SciPy-NMeth}; 
\textsc{Matplotlib} \citep{Hunter:2007}; 
\textsc{Astropy} \citep{2013A&A...558A..33A, 2018AJ....156..123A, 2022ApJ...935..167A}; 
\texttt{EAZY-PY} \citep{brammer2008eazy};
\textsc{Bagpipes} \citep{carnall2018inferring}; 
\textsc{mpi4py} \citep{dalcin2021mpi4py}; 
\textsc{Pickle} \citep{van1995python}.

We also extend our gratitude to Mr. Nicolo Dalmasso and Dr. Nicha Leethochawalit for generously sharing the SFR, sSFR, and stellar mass data for their morphology-detected merging galaxies from \cite{dalmasso2024rate}. These data have been extremely valuable in investigating the impact of mergers on star-forming properties at different merger stages. We also thank Dr. Steven Willner and Dr. Elias Koulouridis for their constructive comments and suggestions on the manuscript.

\section*{Data Availability}
The data underlying this article is made available by \cite{2023MNRAS.518.4755A, 2024ApJ...965..169A, 2023ApJ...952L...7A, austin2024epochs, harvey2024epochs, 2024arXiv240714973C}. The catalogues of the sample discussed herein may be acquired by contacting the corresponding author.



\bibliographystyle{mnras}
\bibliography{main} 



\section*{}
\noindent
\textit{\small{
$^{1}$ Jodrell Bank Centre for Astrophysics, University of Manchester, Oxford Road, Manchester, UK \\
$^{2}$ Kavli Institute for Cosmology, University of Cambridge, Madingley Road, Cambridge CB3 0HA, UK \\
$^{3}$ Cavendish Laboratory – Astrophysics Group, University of Cambridge, 19 JJ Thomson Avenue, Cambridge CB3 0HE, UK \\
$^{4}$ Department of Physics \& Astronomy, University of Victoria, Finnerty Road, Victoria, British Columbia, V8P 1A1, Canada \\
$^{5}$ Institute for Astronomy, University of Edinburgh Royal Observatory, Blackford Hill, Edinburgh, EH9 3HJ, UK \\
$^{6}$ Center for Astrophysics | Harvard \& Smithsonian, 60 Garden Street, Cambridge, MA 02138, USA \\
$^{7}$ School of Earth and Space Exploration, Arizona State University, Tempe, AZ 85287-1404, USA \\
$^{8}$ University of Louisville, Department of Physics and Astronomy, 102 Natural Science Building, Louisville, KY 40292, USA \\
$^{9}$ Department of Physics, University of the Basque Country UPV/EHU, E-48080 Bilbao, Spain \\
$^{10}$ DIPC, Basque Country UPV/EHU, E-48080 San Sebastian, Spain \\ 
$^{11}$ Ikerbasque, Basque Foundation for Science, E-48011 Bilbao, Spain \\
$^{12}$ Space Telescope Science Institute, 3700 San Martin Drive, Baltimore, MD 21218, USA \\
$^{13}$ Association of Universities for Research in Astronomy (AURA) for the European Space Agency (ESA), STScI, Baltimore, MD 21218, USA \\
$^{14}$ Center for Astrophysical Sciences, Department of Physics and Astronomy, The Johns Hopkins University, 3400 N Charles St., Baltimore, MD 21218, USA \\
$^{15}$ Department of Statistical Science, Wake Forest University, Winston-Salem, NC, USA \\
$^{16}$ International Centre for Radio Astronomy Research (ICRAR) and the International Space Centre (ISC), The University of Western Australia, M468, 35 Stirling Highway, Crawley, WA 6009, Australia \\
$^{17}$ Department of Astronomy/Steward Observatory, University of Arizona, 933 N Cherry Ave, Tucson, AZ, 85721-0009, USA \\
$^{18}$ National Research Council of Canada, Herzberg Astronomy \& Astrophysics Research Centre, 5071 West Saanich Road, Victoria, BC V9E 2E7, Canada \\
$^{19}$ ARC Centre of Excellence for All Sky Astrophysics in 3 Dimensions (ASTRO 3D), Australia \\
$^{20}$ INAF-Osservatorio Astronomico di Trieste, Via Bazzoni 2, 34124 Trieste, Italy \\
$^{21}$ Department of Physics and Astronomy, University of Missouri, Columbia, MO 65211, USA}}

\appendix

\section{Merger-Triggered Star Formation Across Different Line-of-Sight Criteria}
\label{sec: appendix 1}
\begin{figure*}
    \centering
    \begin{subfigure}{0.99\linewidth}
        \centering
        \includegraphics[width=\linewidth]{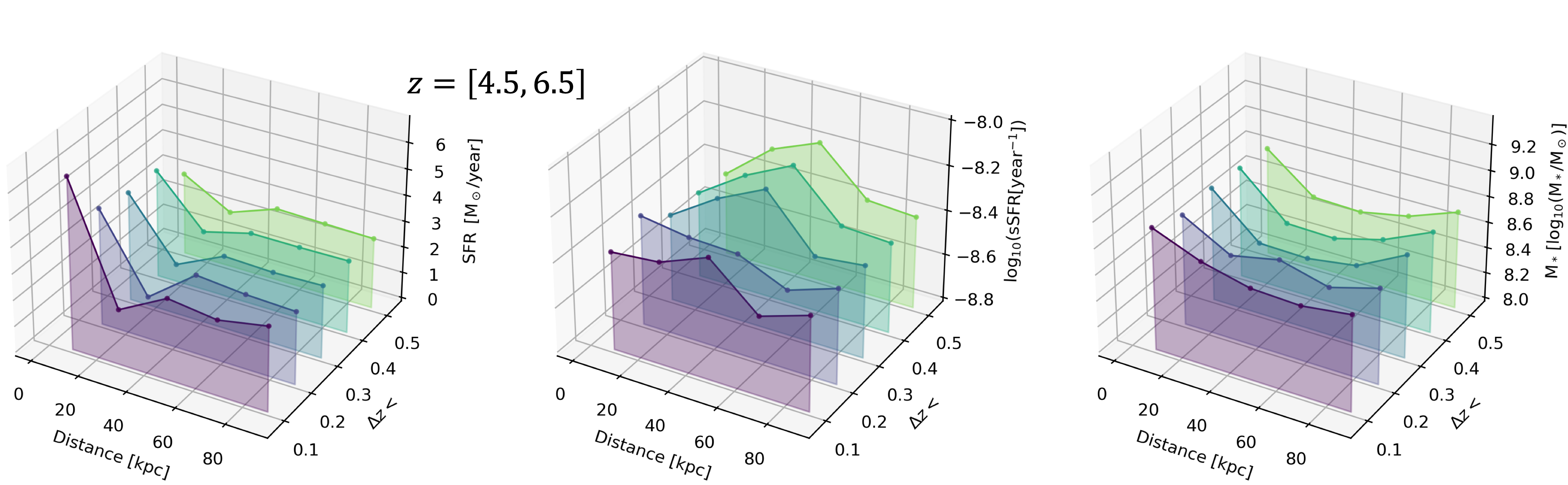}
        \caption{Variation of SFR, sSFR, and stellar mass with projected separation in observational data, using five different $\Delta z$ upper limit criteria at $z = [4.5, 6.5]$.}
        \label{fig: 3d 45 65}
    \end{subfigure}%
    \hfill
    \begin{subfigure}{0.99\linewidth}
        \centering
        \includegraphics[width=\linewidth]{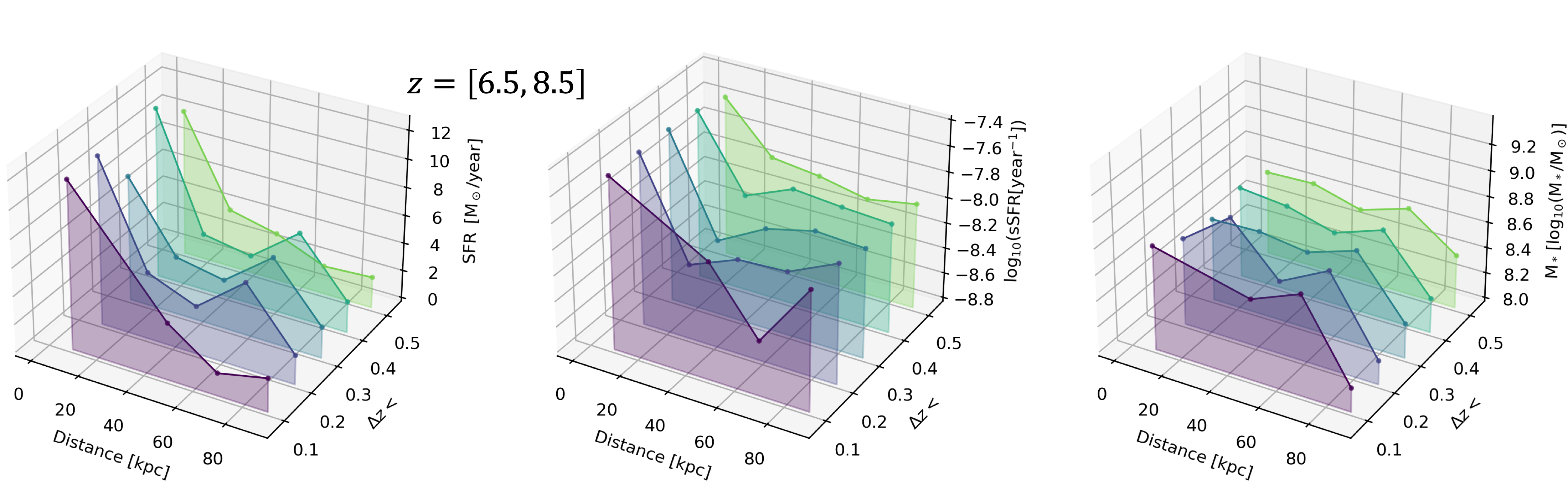}
        \caption{Variation of SFR, sSFR, and stellar mass with projected separation in observational data, using five different $\Delta z$ upper limit criteria at $z = [6.5, 8.5]$.}
        \label{fig: 3d 65 85}
    \end{subfigure}%

    \begin{subfigure}{0.99\linewidth}
        \centering
        \includegraphics[width=\linewidth]{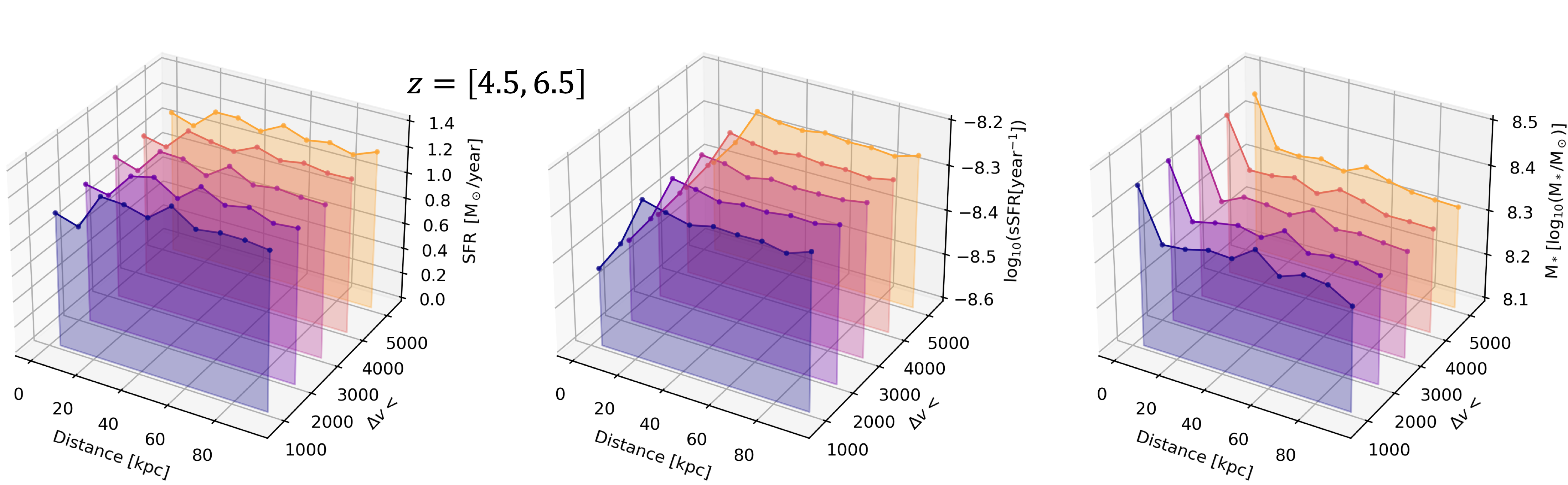}
        \caption{Variation of SFR, sSFR, and stellar mass with projected separation in SC-SAM simulation data, using five different $\Delta v$ upper limit criteria at $z = [4.5, 6.5]$.}
        \label{fig: 3d 45 65 sc-sam}
    \end{subfigure}%
    \hfill
    \begin{subfigure}{0.99\linewidth}
        \centering
        \includegraphics[width=\linewidth]{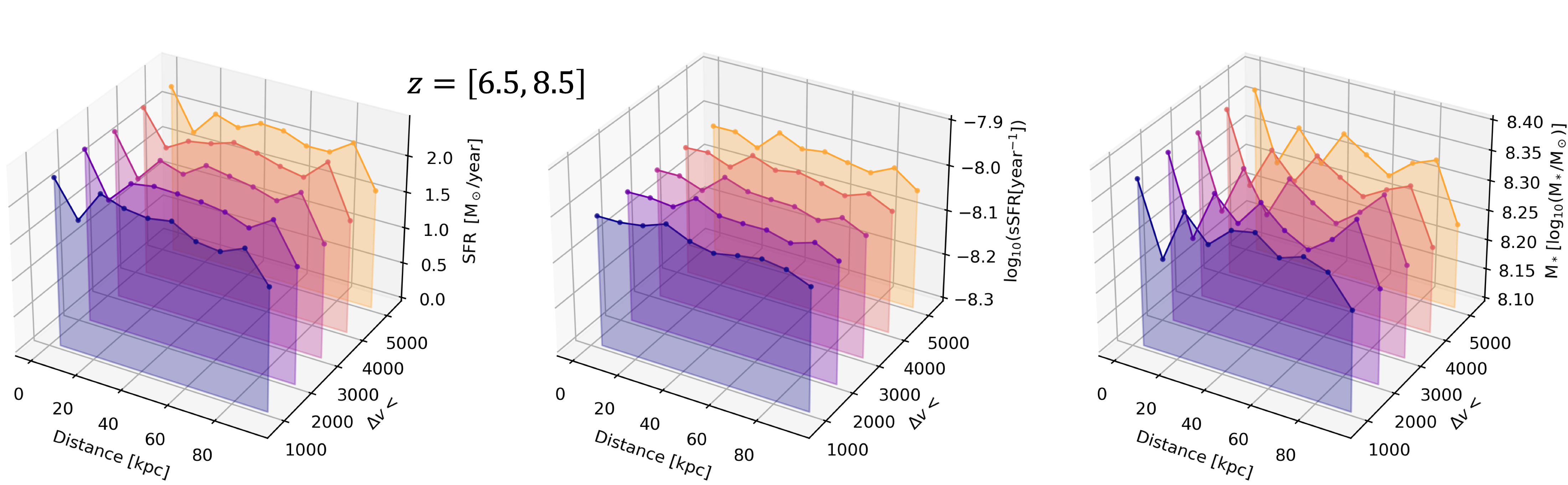}
        \caption{Variation of SFR, sSFR, and stellar mass as a function of projected separation in the SC-SAM simulation data, using five different $\Delta v$ upper limit criteria at $z = [6.5, 8.5]$.}
        \label{fig: 3d 65 85 sc-sam}
    \end{subfigure}%

    \caption{Four figures illustrating the variation of SFR, sSFR, and stellar mass as a function of projected separation in both observational and simulation data, using different line-of-sight selection criteria.}
    \label{fig: 3d 45 65 85 all}
\end{figure*}

In this appendix, we present the evolution of galaxy SFR, sSFR, and stellar mass as a function of projected separation to the closest neighbor, under different line-of-sight selection criteria. We carry out this test as it is possible that an enhanced signal of star formation would be present at some combination of pair separation and velocity/redshift different.   A fiducial $\Delta z < 0.3$ for observational data and $\Delta v < 1000$ km/s for simulation data were chosen, and our results under these criteria have been presented and discussed in detail in \autoref{sec: star_formation_enhancement}. Here, we explore the variation of these properties using five different $\Delta z$ upper limits: 0.1, 0.2, 0.3, 0.4, and 0.5 for the observational data, and five different $\Delta v$ upper limits: 1000, 2000, 3000, 4000, and 5000 km/s for the simulation data, with results presented in \autoref{fig: 3d 45 65 85 all}. We also tested results using $\Delta v$ upper limits of 100, 200, and 300 km/s for the SC-SAM data, and the results are very similar to those obtained with other $\Delta v$ upper limits. Overall, we find that the degree of SFR enhancement from observations at $\Delta z < 0.1$ and $r_p < 20$ kpc is more pronounced than the results at other $\Delta z$ criteria ($\Delta z < 0.2, 0.3, 0.4$). Apart from this, we do not observe significant differences between the results obtained with different line-of-sight criteria. 

\bsp	
\label{lastpage}
\end{document}